\pdfoutput=1
\documentclass[acmsmall,screen,nonacm]{acmart}
\renewcommand\footnotetextcopyrightpermission[1]{} 

\usepackage{algorithm2e}
\usepackage{csquotes}

\usepackage{amsmath,amssymb,amsthm}
\usepackage{thmtools}
\usepackage{thm-restate}
\usepackage{cleveref}
\usepackage[nowarn,acronym]{glossaries}
\usepackage{booktabs}
\usepackage{multirow}
\usepackage{subcaption}
\usepackage{lineno}
\usepackage{pifont}
\usepackage[final]{listings}
\usepackage{wasysym}
\usepackage{wrapfig}
\usepackage{array}
\usepackage{stmaryrd}
\usepackage[inline]{enumitem}
\usepackage{float}
\usepackage{comment}
\usepackage{wrapfig}
\usepackage{tikz}
\usepackage{tikz-qtree}
\usepackage{enumitem}
\usepackage{todonotes}
\usepackage{afterpage}
\usepackage{pdflscape}

\newcommand{\myparagraph}[1]{\medskip \noindent \textbf{#1.}}

\floatstyle{plaintop}

\setlist[description]{labelindent=1em,noitemsep,parsep=0.5ex}

\newcommand{\ie}{\emph{i.\@{}e.\@}\xspace}
\newcommand{\eg}{\emph{e.\@{}g.\@}\xspace}
\newcommand{\vs}{\emph{vs.\@}\xspace}

\newcommand{\etc}{\emph{etc.\@{}}\xspace}

\let\existssymb\exists
\let\forallsymb\forall

\renewcommand{\exists}[1]{\existssymb{#1.}\,}
\renewcommand{\forall}[1]{\forallsymb{#1.}\,}

\newcommand{\states}{\mathcal S}

\newcommand{\step}{\to}

\DeclareMathOperator{\pc}{pc}

\newcommand{\binsecQRSE}{\textsc{Binsec/QRSE}}

\newcommand{\restrictinput}[2]{\left.#1\right|_{#2}}
\newcommand{\restrictpath}[2]{\left.#1\right|^{#2}}
\newcommand{\restrictbound}[2]{\left.#1\right|^{\le #2}}
\newcommand{\reaching}[2]{R\left(#1, #2, O\right)}
\newcommand{\tracesof}[1]{T\left(#1\right)}

\newcommand{\oraclename}{\code{ComputePQR}\xspace}
\newcommand{\oracle}[3]{\oraclename\ensuremath{(#1,#2,#3)}}
\newcommand{\fmset}{\mathcal F}

\newcommand{\varset}{\mathcal{V}}
\newcommand{\bool}{\mathbb{B}}
\newcommand{\uint}{\mathbb{N}}

\newcommand{\true}{\top}
\newcommand{\false}{\bot}
\newcommand{\concat}{||}

\newcommand{\modelsof}[1]{M\left(#1\right)}
\newcommand{\card}[1]{\left|#1\right|}
\newcommand{\varsof}[1]{V\!\left(#1\right)}
\newcommand{\mc}[1]{\sharp{}\left(#1\right)}
\newcommand{\algo}[1]{\textsc{#1}}
\newcommand{\witness}[2]{w_{#1}\left({#2}\right)}
\newcommand{\update}[3]{#1[#2 := #3]}
\DeclareMathOperator{\emajsatop}{emajsat}
\newcommand{\randsymb}{\rotatebox[origin=c]{180}{$\mathsf{R}$}}
\newcommand{\rand}[2][]{\randsymb^{#1}{#2.}\,}
\newcommand{\emajsatfun}[2]{\emajsatop_{#1}\left(#2\right)}

\DeclareMathOperator{\ite}{ite}

\newcommand{\pb}[1]{\textbf{#1}}
\newcommand{\emajsat}{\pb{E-MAJSAT}\xspace}
\newcommand{\femajsat}{$f$\pb{-E-MAJSAT}\xspace}
\newcommand{\sharpsat}{\pb{$\sharp{}$SAT}\xspace}
\newcommand{\instantiate}[2]{\left.{#1}\right|_{#2}}
  \newcommand{\qr}[2]{q\left({#1}, {#2}\right)}

\newcommand{\LTadd}[1]{}
\newcommand{\LTalter}[2]{#1}

\newcommand{\LTinput}[1]{}
\newcommand{\code}[1]{\LTalter{\texttt{#1}}{CODE}}

\allowdisplaybreaks[4]

\author{S\'ebastien Bardin}
\affiliation{\institution{Universit\'e Paris-Saclay, CEA, List}\country{France}}
\email{first.last@cea.fr}
\author{Guillaume Girol}
\affiliation{\institution{Universit\'e Paris-Saclay, CEA, List}\country{France}}
\email{first.last@m4x.org}
\title{A Quantitative Flavour of Robust Reachability}
\begin{document}


\newacronym{ASLR}{ASLR}{Address Space Layout Randomization}
\newacronym{SSP}{SSP}{Stack Smashing Protection}
\newacronym{CTL}{CTL}{Computational Tree Logic}
\newacronym{BMC}{BMC}{Bounded Model Checking}
\newacronym{SE}{SE}{Symbolic Execution}
\newacronym{RSE}{RSE}{Robust Symbolic Execution}
\newacronym{QRSE}{QRSE}{Quantitative Robust Symbolic Execution}
\newacronym{SMT}{SMT}{Satisfiability Modulo Theory}
\newacronym{CFG}{CFG}{Control Flow Graph}
\newacronym{CTF}{CTF}{Capture The Flag}
\newacronym{MBQI}{MBQI}{Model Based Quantifier Instantiation}
\newacronym{NFS}{NFS}{Network File System}
\newacronym{ATL}{ATL}{Alternating-Time Temporal Logic}
\newacronym{cnf}{CNF}{Conjunctive Normal Form}
\newacronym{ddnnf}{d-DNNF}{deterministic Decomposable Negational Normal Form}
\newacronym{dec}{decision-DNNF}{decision Decomposable Negational Normal Form}
\newacronym{counting}{cd-DNNF}{counting decision Decomposable Negational Normal Form}
\newacronym{dag}{DAG}{Directed Acyclic Graph}
\newacronym{aeg}{AEG}{Automatic Exploit Generation}
\newacronym{SSAT}{SSAT}{Stochastic boolean Satisfiability}
\newacronym{QBF}{QBF}{Quantified Boolean Formula}



\begin{abstract}
  Many software analysis techniques attempt to determine whether bugs are
  reachable, but for security purpose this is only part of the story 
   as it does not indicate whether the
  bugs found could be easily triggered by an attacker. 
  The recently introduced notion of robust reachability aims at filling this gap 
  by distinguishing the input controlled by the attacker from those that are not.  
  Yet, this {\it qualitative} notion may be too  strong in practice, leaving apart 
  bugs which are mostly  but not fully  replicable. 
%
  %
  %
  We aim here at proposing a \textit{quantitative} version of  robust reachability, 
  more flexible and still amenable to automation. 
  We propose  \textit{quantitative robustness}, a metric expressing how easily
  an attacker can trigger a bug while taking into account  that he can only
  influence part of the program input, 
  together with a dedicated  quantitative symbolic executon technique (QRSE).   
  %
  %
  %
  Interestingly, QRSE relies on  a variant of model counting  (namely, functional
  E-MAJSAT) unseen so far in formal verification, but which has been studied in 
  AI domains  such as Bayesian network, knowledge representation and  probabilistic planning. 
  %
  %
  %
  Yet, the existing solving methods from these fields turn out to be unsatisfactory for formal verification purpose, 
  leading us to propose a novel parametric method. 
%
  %
  These results have been implemented and evaluated over two security-relevant case studies, 
  allowing to demonstrate the feasibility and relevance of our ideas. 
%
\end{abstract}


\maketitle

\thispagestyle{empty}

\section{Introduction}

\myparagraph{Context \& Problem}
Many software analysis problems are reduced to the reachability of a specific
condition, for example a bug. Yet, for security analysis such as  vulnerability assessment,
reachability is too weak: it proves that the bug exists in at least
one situation, but the security impact depends on further parameters, notably
whether
this situation is unique or  depends on conditions which
are out of reach for the attacker. 
Recent work~\cite{cavRobust} introduced the stronger notion of \textit{robust reachability} to determine
whether an attacker can reproduce a bug reliably: a bug is robustly reachable if
an attacker can choose the part of the program input he
controls so that the bug is triggered, whatever the other input values. 
%

Unfortunately, robust reachability over-compensates the weakness of reachability and ends up too
strong: it requires that when the attacker
plays optimally by choosing the part of input he controls at his advantage, the bug is triggered 100\% of the time. Naturally, we would
also want to detect bugs which  happen 99\% of the time, while still
dismissing those which happen for one input out of $10^{30}$ at best. Yet, currently,  both are
reachable and none  is robustly reachable, hence the need for a more precise notion and appropriate tooling.   
%
%
%

\myparagraph{Goal and challenges} 
We want to provide a quantitative assessment of the ability of the attacker to perform his attack, 
in order to distinguish between unlikely-but-not-zero and  99\% success attacks.  
%
More precisely, we want   a  \textit{quantitative} 
counterpart to robust reachability, like the
non-interference~\cite{goguenNonInterference} community developped
quantitative information
flow~\cite{heusserQuantifyingInformationLeaks2010} to make it less strict, or
the similar shift from model checking to probabilistic model
checking~\cite{azizVerifyingContinuousTime1996}.

This sounds like model counting in the sense
that we count inputs that trigger the bug, but we additionally want to take the presence of
the attacker into account like robust reachability does: attacker input is chosen as worst
case, and other input is counted. In that sense, the underlying counting problem is actually 
very different from those commonly used in quantitative verification, such as (plain) model counting and projected model counting~\cite{existsSATProjected2015}.

 
\myparagraph{Proposal}
We split the program 
input into attacker-controlled input $a$ and uncontrolled input $x$. We define
\emph{quantitative robustness} as
the proportion of uncontrolled inputs $x$ which trigger the bug when the
attacker chooses controlled input $a$ optimally. 
 If
$f$ is a function of $(a, x)$ expressing that the bug is hit, 
%
%
we want $\max_{a} |\{x\mid f(a, x)\}|$,
normalized between 0 and 1.

Starting from this definition, we study the properties of quantitative robustness and propose 
a  bounded-verification  algorithm for this problem, inspired by  symbolic execution. 
Our algorithm relies on the ability to compute path-wise quantitative robustness. While uncommon in formal verification, it turns 
out that for the propositional case (and extensions, such as bitvectors + arrays) this problem as been studied in some  
AI sub-communities  under the name of \femajsat~\cite{emajsat}.    
Unfortunately, the solvers developped there~\cite{complan,complan+,ssatABC,dcssat,maxcount} are
often tuned for other kinds of instances, and for example
some algorithmic improvements developped for probabilistic planning
turn out
detrimental for our purposes. We therefore design a new parametric approximate   algorithm
to
better fit this new domain of application.

\myparagraph{Contributions} We claim the following contributions:
\begin{itemize}

  \item We define a quantitative pendant of robust reachability called
    quantitative robustness (\Cref{sec_p_qr}), which generalize both reachability and robust reachability. We show that quantitative robustness
    has better behavior on branches than robust reachability, allowing incremental path reasoning and removing the need for merging. Interestingly, quantitative robustness is distinct from 
    prior attempts at quantitative program analysis, such as probabilistic model checking or quantitative information flow. We also discuss the relationship with existing quantitative formalisms such as probabilistic temporal logics and games;  




  \item  We propose  \gls{QRSE} (\Cref{sec_p_qrse}), a variant of symbolic execution for computing quantitative robustness, modulo an oracle  for path-wise quantitative robustness.  
  We discuss correctness and completeness issues (includig when the oracle is approximated). Our insights on  the structure of quantitative robustness  bring  interesting properties of QRSE.   
 Notably,  QRSE  does not stricly require path merging, re-establishing the
 symmetry in deduction power  between symbolic execution~\cite{cadarSE} and
 bounded model checking~\cite{cbmc} that is broken in the case   of Robust Symbolic Execution (RSE) for robust reachability.  
This is important as single-path methods such as symbolic execution are considered  more scalable than all-path methods such as bounded model checking;   
%




  \item  We  propose a way to effectively compute path-wise quantitative robustness  when variables range other finite domains (typically, bitvectors and arrays) through a reduction to  \femajsat (\Cref{sec:reduction_counting}), a counting problem studied  
  in some subfields of  AI (Bayesian reasoning, probabilistic planning,
  knowledge representation). To our knowledge, this is the first time that this
  problem is used in a formal verification context -- it is distinct from
  typical  model counting and projected model
  counting~\cite{existsSATProjected2015}. 
  As off-the-shelf methods from AI turn out to be inefficient or imprecise for our purpose, we introduce a novel parametric algorithm for \femajsat, where one can tune
    the trade-off precision \vs  performance by a technique we call
    \emph{relaxation} (\Cref{sec_p_relaxation}). Extreme values of the parameter
    degenerate into already known techniques;

\item We have implemented these ideas in two tools:  BINSEC/QRSE  and the Popcon solver (\Cref{sec_p_evaluation}). First experiments demonstrate the feasibility and relevance of our ideas on medium size examples taken from realistic security contexts (physical fault injection over security devices, and  the analysis of a stack buffer overflow CVE in \code{libvncserver}), as well as the interest of our new solver.  Especially, we show that QRSE enables finer bug triage
    depending on the ability of an attacker to trigger bugs compared to symbolic
    execution and robust symbolic execution, and that for  \femajsat problems arising in
    \gls{QRSE}, relaxation solves more problems than other techniques while
    keeping low approximation.




\end{itemize}

Quantitative robustness is  a new compromise to assess the replicability of a
bug. We believe this is an interesting step toward security-relevant quantitative program analysis. Interestingly, while 
quantitative robustness possibly opens new opportunities for formal methods in security analysis, it also draws  new  connexions with 
notions originating from different AI communities.


\medskip 


\section{Motivating example}\label{sec_p_motivating}
Loosely inspired by CVE-2019-15900 where \code{doas} grants privilege depending on
uninitialized memory, consider in \Cref{fig_p_motivating} the case
of two network servers incorrectly using initial memory to determine the privileges of
clients. 
Whether a client can perform sensitive commands
depends on a \code{privilege\_level} which is accessed through a getter
\code{get\_privilege\_level}. We want to
consider the consequences of a bug this getter incorrectly
returns uninitialized memory modeled as random garbage.

\newcommand{\hl}[1]{{\color{blue}\textbf{\underline{#1}}}}
\begin{figure*}
  \begin{minipage}{0.46\linewidth}
    \centering
    \begin{lstlisting}[basicstyle=\scriptsize\ttfamily,escapechar=@]
/* main privilege levels */
#define DEFAULT_PRIVILEGE_LEVEL 1
#define OPERATOR_LEVEL 100
#define ADMIN_LEVEL 9000
/* commands */
#define DROP_PRIVILEGE 0
#define DROP_PRIVILEGE_LEGACY 1
#define GET_VERSION 2
#define SUDO 3

uint32_t uninit; // random garbage
uint32_t privilege_level = DEFAULT_LEVEL;

void set_privilege_level(uint32_t new) {
  privilege_level = new;
}

uint32_t get_privilege_level() {
  // bug: return uninitialized memory
  return uninit;
}
\end{lstlisting}
\end{minipage}
  \begin{minipage}{0.5\linewidth}
    \begin{lstlisting}[basicstyle=\scriptsize\ttfamily,escapechar=@]
void prog1(uint32_t command, uint32_t argument) {
  if (command == GET_VERSION) {
    /* harmless */
  } else {
    /* command is sudo */
    if (get_privilege_level() == OPERATOR_LEVEL) {
      set_privilege_level(ADMIN_LEVEL);
    }
  }
}

void prog2(uint32_t command, uint32_t argument) {
  switch (command) {
    case GET_VERSION: /*harmless*/ break;
    case DROP_PRIVILEGE: case DROP_PRIVILEGE_LEGACY:
      if (argument<get_privilege_level()) {
        set_privilege_level(argument);
      }
  }
}
\end{lstlisting}
\end{minipage}

\caption{\code{prog1} and \code{prog2} are both vulnerable, but one is more than the other}
\label{fig_p_motivating}
\end{figure*}

We compare two versions of the server: \code{prog1} and \code{prog2},
and we consider a network attacker who can send one request under the
form of a pair \code{command}, \code{argument} passed to either function
\code{prog1} or \code{prog2}. He cannot influence other parameters, notably
uninitialized memory \code{uninit}.
Is it possible that the attacker obtains privilege level greater or equal to
\code{ADMIN\_LEVEL} by submitting a carefully chosen command and argument to these
functions?
For \code{prog1}, this happens when the following formula
\(
  f_1 \triangleq \code{command} \neq 2 \wedge \code{uninit} = 100
\)
is satisfied,
and for \code{prog2} when
\(
  f_2 \triangleq \code{command} \in \{0, 1\} \wedge 9000 \le \code{argument} <
  \code{uninit}
\).
In \code{prog1}, when the attacker plays perfectly by choosing $\code{command} = 
1$, he needs to be lucky: only one value of \code{uninit}
out of $2^{32}$
lets him win. To the contrary,
in \code{prog2}, for $\code{command} = 1$ and $\code{argument} = 9000$, more
than 99\% of
values of \code{uninit} will let the attacker achieve his
goal.
We want to develop an automated machinery to back this intuition.

\myparagraph{Qualitative methods} Traditional bug finding techniques  are of little use here: they prove that the attack
is \emph{reachable}, \ie{} that  formulas $f_{1}$ and $f_{2}$
admit both at least one solution. 
We can refine: robust reachability~\cite{cavRobust} states that
the attack always works when the attacker plays perfectly: $\exists{
\code{command}, \code{argument}} \forall{\code{uninit}} f_{x}$, but in
our case this is too
strict as neither program satisfies it.

\myparagraph{Model counting} Where these \textit{qualitative} techniques fail to distinguish our two programs, maybe 
a more \textit{quantitative} one will bear fruit. For example, we could compare the number
of solutions of $f_{1}$ and $f_{2}$,
or rather their density in a search space of size $2^{96}$.
This is reminiscent of probabilistic symbolic
execution~\cite{geldenhuysProbabilisticSymbolicExecution2012}.
For $f_1$, this density is $\frac{(2^{32}-1)\times 2^{32}}{2^{96}}
\simeq 2.3\cdot10^{-10}$, and for 
$f_2$ it is
$\frac{(2^{32}-9001)(2^{32}-9000)}{2^{96}} \simeq 2.3\cdot10^{-10}$.
These values are very close, and worse, they compare
in order opposite to what we expect: $f_1 > f_2$.

\myparagraph{Our approach} The missing ingredient here is to take into account the threat model: the attacker
will choose the best possible input he can, \ie \texttt{command = 1} and
\code{argument} = 9000, but he
cannot influence the value of \texttt{uninit}. What we want to compute
is the amount of solutions for the value of \texttt{command} and \code{argument} most favorable to
the attacker:
\begin{align}
  \max_{\substack{\code{command}\\ \code{argument}}} |\{\mathtt{uninit} \mid
  f_{1}\}| &= |\{100\}| = 1 \label{eqn_p_popcon_motivating_low}\\
  \max_{\substack{\code{command}\\ \code{argument}}} |\{\mathtt{uninit}
    \label{eqn_p_popcon_motivating_high}\mid
f_{2}\}| &= |[9001; 2^{32}-1]| = 2^{32} - 9001
\end{align}
These numbers can be fairly compared as the search space has the same size
($2^{32}$) but in the general case we will consider a proportion of inputs
instead, which we call \emph{quantitative robustness}. Quantitative robustness 
does align to the intuition we had: it is low ($2.3\cdot10^{-10}$) for
\code{prog1} but very close to 1 for \code{prog2}\footnote{Approximately 0.9999979043.}. 

The problem of doing computations like
\cref{eqn_p_popcon_motivating_low,eqn_p_popcon_motivating_high} on a boolean formula is known as functional
\emajsat~\cite{emajsat}, or \femajsat for short. Solvers exist for this problem
but,
although some of them~\cite{dcssat,complan} can obtain
\cref{eqn_p_popcon_motivating_low} in few seconds, we know of no solver able to
obtain \cref{eqn_p_popcon_motivating_high} even at the price of reasonable
approximation.
Taking inspiration from existing knowledge-compilation based algorithms, we
propose a new technique called relaxation that offers an interesting
trade-off between performance and precision.
For \code{prog2} we obtain (with parameter BFS(40)) in about 1 second that the quantitative robustness of privilege
escalation is
comprised between 0.9963 and 1.
This is enough to conclude that there are many more initial states that let the attacker exploit the vulnerability in \code{prog2} than in
\code{prog1}. We interpret this as a sign that this bug is presumably more
severe in \code{prog2} than in \code{prog1}.

\myparagraph{Summary}
Qualitative techniques based on reachability and robust reachability cannot
distinguish \code{prog1} from \code{prog2}, whereas in practice an attacker has
many more opportunities to trigger the bug in \code{prog2}. Quantitative
robustness clearly discriminates between the two, but this is not only because
it is quantitative. Compared to probabilistic symbolic
execution~\cite{geldenhuysProbabilisticSymbolicExecution2012}, quantitative
robustness better fits security contexts by using a variant of model
counting which can distinguish between attacker-controlled inputs and
uncontrolled inputs.

\myparagraph{Remark} 
We are counting models without assigning a weight, or rather a probability, to
each of them. This amounts to  assigning a uniform
distribution to uncontrolled inputs. 
We discuss this point in \Cref{sec_p_qr_def}.

\section{Background}
\label{sec_p_background}

A program $P$ is represented a transition system with
transition relation $\step$ over the set of states $\states$. A trace is a
succession of states respecting $\step$; the set of traces of a program $P$ is
$\tracesof{P}$. Each state has a corresponding location in the source of the
program, a \emph{path} is a succession of locations. The first state of the program is determined by the input $y$ of
the program; we assume a deterministic program whose randomness is due to input.
$\restrictinput{P}{y}$ is the program identical to $P$ but executed
on input $y$. We adopt the threat model of \citet{cavRobust}: input $y$ is a
pair $(a, x)$ of \emph{controlled inputs} $a$ chosen by the attacker in a set
$\mathcal A$, and
\emph{uncontrolled inputs} $x\in\mathcal X$ unknown to the attacker and
uninfluenced by him.

\paragraph{Reachability, robust reachability} For $O$ a set of finite traces, we
say that $O$ is \emph{reachable} in $P$ when $\tracesof{P} \cap O \neq \varnothing$, meaning that $P$
admits a
trace reaching the goal, and that $O$ is 
\emph{robustly reachable}~\cite{cavRobust} when
$\exists{a\in \mathcal A} \forall{x \in\mathcal X} \tracesof{\restrictinput{P}{(a, x)}} \cap O \neq
\varnothing$, meaning that for some controlled input $a$, for all uncontrolled
inputs $x$, the target is reached.

    \RestyleAlgo{boxed}
\begin{wrapfigure}[11]{r}{0.4\textwidth}
  \begin{algorithm}[H]
      \LinesNumbered
      \DontPrintSemicolon
      \KwData{bound $k$, target $O$}
      \SetKwFunction{GetPaths}{GetPaths}
      \SetKwFunction{GetPredicate}{GetPredicate}
      \For{path $\pi$ in \GetPaths$(k)$}{
          $\phi := \GetPredicate(\pi, O)$\;
          \lIf{$\exists{a, x}\phi$}{
            \textbf{return} true
        }
      }
      \textbf{return} false\;
      \caption{Reachability of $O$ by symbolic execution}
      \label{alg_p_se}
    \end{algorithm}
  \end{wrapfigure}
\paragraph{Symbolic execution} Reachability can be proved by \glsentryfull{SE}
\cite{cadarSE}. 
SE enumerates all paths $\pi$, converts them to a SMT formula $\pc_\pi^O(a, x)$
called \emph{path constraint} expressing what input $(a, x)$ make the program go
along $\pi$ and reach the goal $O$, and checks whether this formula is
satisfiable. If this is the case, then $O$ is reachable. SE is
correct (detected targets are reachable) and $k$-complete (when bounding paths
to length $k$, a reachable is detected).

Robust Symbolic Execution \gls{RSE}~\cite{cavRobust} proves robust reachability by replacing
satisfiability tests $\exists{a, x}
\pc_\pi^O(a, x)$ in SE by $\exists a \forall x \pc_\pi^O(a, x)$. It is
correct, but not $k$-complete. For $k$-completeness, path
merging~\cite{hansenStateJoiningSplitting2009} is required: 
paths constraints of paths are merged together as $\bigvee_i \pc_{\pi_i}^O(a, x)$.

\section{Quantitative robustness}\label{sec_p_qr}
In this section, we define quantitative robustness and study its behavior along
program paths. 
\subsection{Threat model}
We consider the program as a deterministic system where all sources of
randomness are modeled as explicit inputs. Inputs to the program are partitioned into
\emph{controlled inputs}, chosen by the attacker, and \emph{uncontrolled input},
unknown to the attacker. This threat model is the same as robust reachability~\cite{cavRobust}, 
and it is well adapted to an attacker submitting a request to a non-interactive system
(for example a network server). The request is then a controlled input, and all
other inputs, notably implicit ones like initial memory or randomness, are
uncontrolled. However, this threat model excludes interactive systems, which
is important to keep proof methods tractable.

\subsection{Formal definition}\label{sec_p_qr_def}

  Quantitative robustness is the maximal proportion of uncontrolled inputs that reaches
  the target, for the best controlled input. In anticipation of the needs of computation
  techniques in the next section, we assume that uncontrolled inputs are in finite number.
\begin{definition}[Quantitative robustness]
  We consider the reachability problem associated to program $P$ and target set
  of paths $O$.
  The associated quantitative robustness is
  \[
    \qr{P}{O} \triangleq \frac 1 {|\mathcal X|} \max_{a\in \mathcal A} \left|
    \left\{ x \in \mathcal X \mid
    \tracesof{\restrictinput{P}{(a, x)}} \cap O \neq \varnothing \right\} \right|
  \]
\end{definition}

Extreme values of quantitative robustness correspond to already known properties:
\begin{proposition}
  Quantitative robustness is 0 if and only if the target is not reachable.
  Quantitative robustness is 1 if and only if the target is robustly reachable.
\end{proposition}

Quantitative robustness is designed to
detect
bugs which are nearly robust, but not exactly because for few uncontrolled
inputs the target is missed: they should have a quantitative robustness close to
1.

\myparagraph{Scope \& limitations}
This definition inherits limitations of robust reachability.
The attacker can only submit one input to the system, in one go, and
without knowledge of uncontrolled inputs. While  already covering a wide spectrum of real attacks, this definition forbids interactive
systems. A definition accepting interactive systems is possible but less
tractable.
%
In the same vein, we limit our discussion to the reachability of a (possibly infinite) set of finite traces, which already encompasses
critical scenarios such as buffer and stack  overflows, use-after-free, control-flow hijacking, etc. More advanced properties such 
as hyperproperties (e.g., secret leakages) or infinite traces (e.g., denial of service) are left as future work. 

Model counting brings additional constraints:
 inputs are assumed to be \emph{in finite
number} and \emph{uniformly distributed}.
A straightforward solution to both problems is to consider  the maximal  probability of uncontrolled
input to reach the target, with some
 probability measure over the possibly infinite set $\mathcal X$. Actually, results from \Cref{sec_p_qr,sec_p_qrse} 
should hold in this setting. Yet, we will be left with the problem of designing solvers for the underlying probability estimation 
problem, which does not exist for the moment, to the best of our knowledge.  

Going deeper,  let us argue that these limitations are actually not that much a problem in practice.  
First (finiteness), the theory of arrays + bitectors + uninterpreted functions  is intensively used in security-related program analysis, 
and it has indeed a finite interpretation. 
Second (distribution), while  specifying arbitrary non-uniform input distribution may  seems handy at first, in practice 
determining the probability distribution of uncontrolled inputs is far from
trivial (ex: distribution of system calls such as \code{malloc}), except for a few cases where 
the distribution is specifically intended to be uniform (stack canaries, ASLR influences documented bits, 
or hash function).

\subsection{Quantitative robustness and paths}
Robust reachability can be lost at a branch depending on uncontrolled
input and recovered later when paths meet again. This forces us to merge paths
together. On the other hand, quantitative robustness is not fully lost when
paths separate. We denote the restriction of
  $P$ to paths $\pi_1, \dots, \pi_n$ as $\restrictpath{P}{\pi_1,\dots,\pi_n}$,
  and we start with some properties of quantitative robustness of such a
  restriction.

  \begin{proposition}[Monotonicity of quantitative robustness of paths]
    \label{prop_p_qr_monotonicity_path}
    Let $\pi$ be a path in a program $P$.
    $\qr{\restrictpath{P}{\pi}}{O}\le\qr{P}{O}$.
  \end{proposition}
  \begin{proof} Let $\reaching{P}{a} \triangleq\left\{x \in \mathcal X \mid
    \tracesof{\restrictinput{\restrictpath{P}{\pi}}{(a,x)}} \cap O \neq
    \varnothing\right\}$.\\
    Then: $\qr{P}{O} = \max_a |\reaching{P}{a}| / |\mathcal X|$.
    The result follows from the fact that  
    $\forall{a\in \mathcal A}
     \reaching{\restrictpath{P}{\pi}}{a} \subseteq \reaching{P}{a}$.
\end{proof}
\begin{proposition}[Quantitative robustness of merged paths]
    \label{prop_p_qr_triangular_ineq}
  Let $\pi, \pi'$ be two paths in a program $P$. Then
  \[
    \qr{\restrictpath{P}{\pi,\pi'}}{O}\le
    \qr{\restrictpath{P}{\pi}}{O}+\qr{\restrictpath{P}{\pi'}}{O}
  \]
\end{proposition}
\begin{proof}
  Let $a$ reaching the $\max$ in the definition of
  $\qr{\restrictpath{P}{\pi,\pi'}}{O}$.
  \begin{equation}
  \reaching{\restrictpath{P}{\pi, \pi'}}{a} = \reaching{\restrictpath{P}{\pi
 }}{a} \cup \reaching{\restrictpath{P}{\pi'}}{a}
 \label{eqn_union}
\end{equation}
In terms of cardinal $\left|\reaching{\restrictpath{P}{\pi, \pi'}}{a}\right| = |\mathcal
X|\qr{\restrictpath{P}{\pi, \pi'}}{O}$ by definition of $a$ and
$\left|\reaching{\restrictpath{P}{\pi}}{a}\right|
\le |\mathcal X|\qr{\restrictpath{P}{\pi}}{O}$ by definition of quantitative robustness.
The result follows from a union bound on \cref{eqn_union}.
\end{proof}

Quantitative robustness cannot vanish at a
branch:
\begin{proposition}[Quantitative robustness pseudo-conservation]
  \label{prop_p_qr_conservation}
  Let $\pi_1, \dots, \pi_n$ be paths in a program $P$.
  There exists $1 \le i \le n$ such that $\qr{\restrictpath{P}{\pi_i}}{O} \ge \frac
  1 n \qr{\restrictpath{P}{\pi_1, \dots, \pi_n}}{O}$.
\end{proposition}
\begin{proof}
By contradiction, if $\qr{\restrictpath{P}{\pi_i}}{O}< \frac
1 n \qr{\restrictpath{P}{\pi_1, \dots, \pi_n}}{O}$ for all $i$ from 1 to $n$, then by
\Cref{prop_p_qr_triangular_ineq},  $\qr{\restrictpath{P}{\pi_1, \dots,
\pi_n}}{O} < n \times \frac 1 n \qr{\restrictpath{P}{\pi_1, \dots,
\pi_n}}{O}$ which is absurd.
\end{proof}

\begin{wrapfigure}[12]{r}{0.3\textwidth}
      \centering
      \begin{lstlisting}
void main(a, x) {
  if (x) x++; // $\pi_1$
  else x--;   // $\pi_2$
  if (!a) bug();
}
      \end{lstlisting}
    \caption{An example  where path merging is required in RSE (taken from
    \citet{cavRobust})}
    \label{fig_r_merge_required}
  \end{wrapfigure}
To illustrate why this is good news, consider the case that justified the
necessity of path merging in RSE: \Cref{fig_r_merge_required}. The program $P$ has two paths $\pi$ and $\pi'$
starting at location $s$, selected
depending on an uncontrolled boolean input $x$, and which join again in location
$\ell$. Neither $\pi_1$ nor $\pi_2$ satisfies single path robust reachability, but
$\ell$ is robustly reachable. Robust reachability can \enquote{reappear} from
non-robust paths quite unpredictably, so we are forced to merge all paths to
keep completeness. This is not the case with quantitative reachability as
\Cref{prop_p_qr_conservation} guarantee that one of $\pi_1$ or $\pi_2$ has
quantitative reachability at least $\frac 1 2$.
In this situation one can thus
still detect $\ell$ without path merging by lowering our detection threshold by
one half.

\subsection{Comparison to other quantitative formalisms}
Several domains in software analysis have moved to quantitative approaches for better precision.

\paragraph{Probabilistic reachability}
Program verification is usually encoded as the reachability of an undesirable
condition, so it is natural to consider the probability of reaching it.
For example probabilistic symbolic
execution~\cite{geldenhuysProbabilisticSymbolicExecution2012} attempts to
compute the probability\footnote{Actually, they compute model counts and
therefore assume uniformly distributed inputs, like we do.} of each path, and shows experimentally that one can find
bugs by focusing human analysis on improbable paths. The main difference with
our work is that they compute the probability of a bug happening in a neutral
environment, whereas we take into account the presence of an attacker.

\paragraph{Probabilistic temporal logics} 
Probabilistic logics developped for model checking like
pCTL~\cite{hanssonLogicReasoningTime1994} use Markov chains
instead of model counting on constraints systems. They can express the
probability of complex events in interactive systems with several
rounds of input,
but not systems where two actors have different interests.
Mapping the CTL encoding of robust reachability ($\mathbf{EXAF}\varphi)$ to pCTL
expresses the probability of reaching for a specific attacker whose probability
transition tables are known. This does not fit our use case, where attacker actions should be taken as
worst case and are not known \emph{a priori}.
More expressive logics like
MTL\textsubscript{2}~\cite{jamrogaTemporalLogicStochastic2008}, a generalisation
of ATL~\cite{alurAlternatingtimeTemporalLogic2002}, can
express a worst-case attacker, but they are so general that they lack tractable proof
methods.


\paragraph{Quantitative information flow}
Quantitative information flow attempts to quantify the amount of information that an
attacker can deduce from the observable  behavior of a
system, interpreted as leakage of information. 
The attacker chooses public input to a system, the defenders chose secret inputs,
and the attacker attempts to deduce the secret from the public output.
A central notion is the capacity of the leakage channel: the
logarithm of the number of
public outputs $z$ such that there exists a
pair of (public, private) inputs leading to $z$. This problem is called
\emph{projected model counting}~\cite{existsSATProjected2015} and is distinct
from our approach based on \femajsat.

\section{Quantitative robust symbolic execution}\label{sec_p_qrse}

In this section, we design a method to enumerate all locations with quantitative
robustness above a threshold $Q$, and to know their quantitative robustness,
\eg{} to sort them from most to least robustly
reachable.

Like symbolic execution determines reachability from path-wise reasoning on the
satisfiability, we assume that we can compute quantitative robustness path-wise:
given the program $P$ and target $O$, we have an oracle \oracle{P}{\pi}{O}
which can compute the Path-wise Quantitative Robustness $\qr{\restrictpath{P}{\pi}}{O}$ of
any path $\pi$.

\subsection{Going quantitative from RSE}
We adapt \gls{RSE}~\cite{cavRobust} to this goal by replacing the universal
satisfiability test $\exists a \forall x \pc_\pi^O(a, x)$ by a new test
expressing that many inputs $x$ make $\pc$ true for the best value of $a$.

By replacing universal satisfiability tests by tests that
$\oracle{P}{\pi}{O}$ is greater than the threshold $Q$, we can
enumerate paths which reach the goal with quantitative robustness above $Q$, and
print the computed quantitative robustness for the user. We call this technique
\glsentryfull{QRSE}. More specifically, operating this substitution on RSE
yields QRSE (\Cref{alg_p_qrse}) and on RSE+ (RSE plus path merging) it yields
QRSE+ (QRSE plus path merging, \Cref{alg_p_qrse_merge}).
%

\begin{figure}
  \begin{minipage}{0.48\textwidth}
    \RestyleAlgo{boxed}
    \begin{algorithm}[H]
      \LinesNumbered
      \DontPrintSemicolon
      \KwData{bound $k$, target $O$, threshold $Q$}
      \SetKwFunction{GetPaths}{GetPaths}
      \SetKwFunction{GetPredicate}{GetPredicate}
      $\phi := \bot$\\
      \For{path $\pi$ in \GetPaths$(k)$}{
          $\phi := \GetPredicate(\pi, O)$\;
          $\chi := \oracle{P}{\pi}{O}$\;
          \uIf{$\chi \ge Q$}{
            \tcc{\small{$O$ has quantitative robustness $\ge \chi$}}
            \textbf{return} (true, $\chi$)\;
        }
      }
      \textbf{return} false\;
      \caption{QRSE: Quantitative Robust \gls{SE}}
      \label{alg_p_qrse}
    \end{algorithm}
\end{minipage}
  \begin{minipage}{0.48\textwidth}
    \RestyleAlgo{boxed}
    \begin{algorithm}[H]
      \LinesNumbered
      \DontPrintSemicolon
      \KwData{bound $k$, target $O$, threshold $Q$}
      \SetKwFunction{GetPaths}{GetPaths}
      \SetKwFunction{GetPredicate}{GetPredicate}
      $\phi := \bot$\\
      \For{path $\pi$ in \GetPaths$(k)$}{
          $\phi := \phi \vee \GetPredicate(\pi, O)$\\
          $\chi := \oracle{P}{\pi}{O}$\;
          \uIf{$\chi \ge Q$}{
            \tcc{\small{$O$ has quantitative robustness $\ge \chi$}}
            \textbf{return} (true, $\chi$)\;
        }
      }
      \textbf{return} false\;
      \caption{QRSE+: QRSE with path merging}
      \label{alg_p_qrse_merge}
    \end{algorithm}
  \end{minipage}
\end{figure}

\begin{proposition}[Correctness of QRSE]
  If QRSE reports a target $O$ with quantitative robustness $\chi$, then
  $\qr{P}{O}\ge \chi$.
\end{proposition}
\begin{proof}
  QRSE reaching $O$ proves that there is a path $\pi$ such that
  $\qr{\restrictpath{P}{\pi}}{O} = \chi$. By
  \Cref{prop_p_qr_monotonicity_path}, $\qr{P}{O}\ge \chi$.
\end{proof}

\begin{proposition}[$k$-completeness of QRSE+]
  We remind the reader that we suppose that the domain of inputs is finite.
  $\restrictbound{P}{k}$ denotes the restriction of program $P$ to traces of
  length at most $k$. Let $Q$ be a threshold.
  Assuming solver termination, if a target $O$ has quantitative robustness
  $\qr{\restrictbound{P}{k}}{O} \ge Q$, then it is
  reported by QRSE+ with a quantitative robustness between $Q$ and $\qr{\restrictbound{P}{k}}{O}$.
\end{proposition}
\begin{proof}
  In $\restrictbound{P}{k}$, for each possible input, there is at most one
  maximal path of length at most $k$ (and
  all its prefixes).
  When QRSE+ has explored all
  paths, the path constraint will be equivalent to reaching $O$.
  The oracle on the merged path constraint of all those paths
  will therefore return the desired value $\qr{\restrictbound{P}{k}}{O}$.
  If some subset of these paths has quantitative robustness between $Q$ and
  $\qr{\restrictbound{P}{k}}{O}$, QRSE+ may return early.
\end{proof}

  \myparagraph{Approximations} If we can only approximate $\qr{\restrictpath{P}{\pi}}{O}$ in
  \Cref{prop_p_qrse_emajsat}, we still keep some guarantees: with a lower
  bound QRSE is still correct and with an upper bound QRSE+ is still
  $k$-complete. 


\subsection{Path merging}\label{sec_p_merging} \gls{RSE} requires path merging
for $k$-completeness~\cite{cavRobust}. We want to avoid it for two main reasons: firstly, some paths can be hard to
execute symbolically (\eg because they contain exotic system calls, or dynamic
jumps, \etc), and secondly, merged path constraints are more complex and harder
to solve.
In the quantitative case, we can show that QRSE without path merging is actually
as complete as QRSE with path merging under a reasonable assumption.


\begin{definition}[Badly scaling path merging assumption]
  We assume that merged paths constraints are more difficult to solve than their
  constituents, and
  that there is an integer $\kappa$ such that, when merging the paths
  constraints of more than $\kappa$ paths together, the resulting path
  constraint is so large and/or complex that our implementation of the oracle
  \oraclename will return UNKNOWN.
\end{definition}


\begin{proposition}[QRSE vs QRSE+]
  Under the badly scaling path merging assumption, all locations reported by
  QRSE+ as having quantitative robustness above the threshold $Q$ are also
  reported by QRSE with the threshold $Q/\kappa$.
\end{proposition}
\begin{proof}
  Let $O$ be a target reported by QRSE+ with threshold $Q$. By the badly scaling
  path merging assumption, there are paths $\pi_1, \dots, \pi_n$ with $n\le
  \kappa$ \emph{s.t.} the oracle can compute $\chi \triangleq
  \oracle{P}{\pi_1, \dots, \pi_n}{O}$ with $\chi \ge Q$. By
  \Cref{prop_p_qr_conservation}, there is a path $\pi_i$ such that
  $\qr{\restrictpath{P}{\pi_i}}{O} \ge Q/n \ge Q / \kappa$. As we assume that 
  merged path
  constraints are harder to solve than the original ones, the oracle can compute
  $\qr{\restrictpath{P}{\pi_i}}{O}$ and QRSE
  detects $O$ by path $\pi_i$ with the threshold $ Q / \kappa$.
\end{proof}

In practice, this means that if path merging turns out to be a problem for QRSE+ with
threshold $Q$, then one can run QRSE with threshold $Q/\kappa$ and have the
guarantee of finding
  all targets with quantitative robustness above $Q$ but
  no targets with quantitative robustness below $Q/\kappa$.
The second point ensures we keep a good signal-to-noise ratio.
This principle will be illustrated in our second case study about
\code{libvncserver}
(\Cref{sec_p_bench}).


\section{Path-wise quantitative robustness as a counting problem}\label{sec:reduction_counting}
We now propose an implementation of the oracle for path-wise
quantitative robustness \oraclename required for \gls{QRSE}. We reduce it to a variant of model
counting called \femajsat.

\subsection{Preliminary: the \femajsat problem}\label{sec_p_background_femajsat}

The set $\fmset{}$ of propositional formulas is defined starting from variables
$v\in \varset{}$, and for $f, g \in \fmset{}$ adding negation $\neg f$,
conjunction $f\wedge g$ and disjunction $f\vee g$.
We denote as $\varsof{f}$ the set of variables appearing
effectively in a formula $f$.
Propositional formulas are usually given in \gls{cnf}. A literal is $v$ or $\neg v$
where $v$ is a variable. A clause is a set of literals, interpreted as their
disjunction, and a formula in \gls{cnf} is a set of clauses, interpreted as
their conjunction.

A partial valuation is a partial mapping from a subset of $\varset{}$ to the set $\bool \triangleq \{\true,
\false\}$. One can apply a partial valuation $m$ to a full formula $f$:
$\instantiate{f}{m}$ is the
formula identical to $f$ where variables $v$ in the domain of $m$ are replaced by
$m(v)$. 
 For example, for $f = v_1 \wedge (\neg v_1 \vee v_2)$ and $m = \{v_1 \mapsto
 \true\}$, the formula obtained by applying $m$ on $f$ is $\instantiate{f}{m} = v_2$.
A valuation is complete for $f$ when its domain contains $\varsof{f}$, \ie it associates all
variables to a boolean value. Such a valuation maps a propositional formula to $\bool$
as well.

A complete valuation $m$ is said to be a model of a formula $f$ if
$\instantiate{f}{m} = \true$.
We denote as $\modelsof{f} \triangleq \{ m \in \bool^{\varsof{f}} \mid
  \instantiate{f}{m} =
\true \}$ the set of models of a formula $f$, and as $\mc{f} \triangleq
\card{\modelsof{f}}$ its cardinal.
For example, the models of $v_1 \wedge (v_2 \vee \neg v_2)$ are $\{v_1 \mapsto
  \true, v_2 \mapsto \false\}$ and $\{v_1 \mapsto
  \true, v_2 \mapsto \true\}$.
%
%
Note that this definition depends on the number of variables of a formula.
Therefore, $\mc{v_1} = 1$ whereas $\mc{v_1 \wedge (v_2 \vee \neg v_2)} = 2$.
The literature usually solves this with the notion of smoothness (see below).


\begin{restatable}[\femajsat~\cite{emajsat}]{definition}{deffemajsat}\label{def_p_femajsat}
  \femajsat is the following function
  problem:
     Given a formula $f$ in \gls{cnf} with a partition of variables in $A$
      and $X$: $\varsof{f} = A \uplus X$,
    output  $\displaystyle \emajsatfun{A}{f} \triangleq \max_{a_1, \dots, a_n \in \bool^A} \mc{\instantiate{f}{a_1, \dots,
      a_n}}$.
\end{restatable}

As usual with functional problems, there is a companion decision problem called
\emajsat which tests whether \femajsat is above $2^{|X|-1}$ (or another threshold).
Variables in $A$ are called \emph{choice variables} and variables in $X$ are
called \emph{chance variables}.
The distinction between chance and choice variables the key to
encode the presence of the attacker and the partition of inputs into controlled
and uncontrolled inputs.
\femajsat reduces to \pb{SAT} when $X = \varnothing$ and to \sharpsat when $A =
\varnothing$, so it is at least as hard as these problems.
\emajsat is $\mathrm{NP}^\mathrm{PP}$-complete~\cite{emajsat}, meaning that it would become
NP with a PP oracle.
\subsection{Path-wise quantitative robustness}

We assume path-constraints generated by \gls{SE} are
propositional formulas.
Inputs
are represented as boolean variables: $a\triangleq(a_1, \dots, a_n)$ and
$x\triangleq(x_1, \dots, x_m)$. 
We add two  formulas $h_a(a)$ and
$h_x(x)$ specifying
valid inputs: $\mc{h_a} = |\mathcal{A}|$ and $\mc{h_x} = |\mathcal{X}|$.
$h_a$ and $h_x$ can also be used to express the effect of \code{assume} statements in
the analyzed program.
\begin{proposition}
  \label{prop_p_qrse_emajsat}
  For a path constraint $\pc_\pi^O$ expressed as a
  propositional formula, path-wise quantitative robustness can be reduced to
  \femajsat as follows:
  \[
  \oracle{P}{\pi}{O} =  \emajsatfun{a}{h_a(a) \wedge h_x(x) \wedge \pc_\pi^O(a,
  x)} / {\mc{h_x}}
\]
\end{proposition}

This observation allows implementing \gls{QRSE} presented in
\Cref{sec_p_qrse} with a \femajsat solver. 
\subsection{Beyond SAT}
One of the keys to the success of \gls{SE} is the expressivity of theories
supported by SMT solvers, compared to manual SAT encoding. It is possible to
reduce some (essentially finite) theories to SAT and thus
\Cref{prop_p_qrse_emajsat} by bitblasting. For each model of a SMT formula,
there is a unique corresponding model in the corresponding bitblasted
propositional formula. This guarantees that model counts are preserved during
bitblasting.

For example in our experiments we will focus on the theory of arrays and
bitvectors. Arrays can be eliminated by eager application of the read-over-write
axiom of the theory, and bitvectors can be bitblasted by mimicking the logical
gates used in processors.


\section{Efficient approximation of \femajsat}\label{sec_p_ddnnf}
In this section we turn to the problem of solving \femajsat on a bitblasted path
constraint obtained during \gls{QRSE}. 
As quantitative robustness is only a hint for one dimension of exploitability,
approximate solutions are acceptable, but efficiency is a must.

\subsection{Prior work:  solving \femajsat with decision-DNNF normal form} 
In this section we present one particular kind of techniques to solve \femajsat, based on
a normal form called \gls{dec}~\cite{decisionDNNF}.

\begin{definition}[\glsentrytext{dec}]
  A formula in \gls{dec} is a DAG of the following nodes:
\begin{description}
  \item[True and False nodes] $\true$ and $\false$;
  \item[Decomposable And node] $\bigwedge_{i=1}^n f_i$, where for ${1 \le i, j
    \le n}$, $\varsof{f_i} \cap \varsof{f_j} = \varnothing$, and the children $(f_i)_{1\le i
    \le n}$ are in \gls{dec};
  \item[Decision (or Ite) node] $\ite(v, f, g)$, where $f$ and $g$ denote formulas in
    \gls{dec}, $v$ a variable, and $v \not\in \varsof{f}$, $v \not \in \varsof{g}$.
    If additionally $\varsof{f} = \varsof{g}$ then the formula is said to be
    \emph{smooth}.
\end{description}
\end{definition}
An example is given in \Cref{fig_p_example_formula_relax}. 
$\ite(v, f, g)$ is a shorthand for \enquote{if $v$ then $f$ else $g$}. 
By convention, $\varsof{\true} = \varsof{\false} = \varnothing$, $\mc{\true} =
1$, $\mc{\false} = 0$. This definition is slightly non-standard: literals are
normally included, but we replace
$v$ by $\ite(v, \true, \false)$ and $\neg v$ by $\ite(v, \false, \true)$.
For smooth Ite nodes, we have $\mc{\ite(v, f, g)} = \mc{f} + \mc{g}$. Without
smoothness, one must reason about pairs $(\mc{f}, \varsof{f})$ instead of $\mc{f}$
which makes the formal treatment considerably heavier.
As usual in the literature, we present the formalism on smooth formulas
only, which can be done without loss of
generality~\cite{darwicheTractableCountingTheory2000} as a formula can be made
smooth in polynomial time.

\paragraph{Compilation} Model counting of a formula in \gls{dec} can be done in linear
time~\cite{darwicheDecomposableNegationNormal2001} (the algorithm is a
special case of \Cref{def_p_emajsat_compute}). This reduces
model counting to the process of converting a CNF formula  to an   equivalent
\gls{dec} formula, which is called
\emph{compilation}. D4~\cite{d4} is a \gls{dec} compiler.
Compilers for a looser normal form called
\gls{ddnnf}~\cite{darwicheDecomposableNegationNormal2001} are more common,
but
interestingly, while \gls{ddnnf} compilers like C2D~\cite{c2d} and Dsharp~\cite{dsharp} officially
output \gls{ddnnf},  they actually produce the stricter \gls{dec}.
All formulas can equally be encoded in either normal forms, so w.l.o.g we
present all algorithms for \gls{dec}.
Compilation is significantly more expensive than model counting on the resulting
\gls{dec} formula: about 96\% of runtime on our test suite of
\Cref{sec_p_bench}.

\newcommand{\mynode}[2]{#1:\ \textcolor{red}{#2}}
\newcommand{\literal}[2]{[.{\mynode{$\ite(#1)$}{#2}} {\mynode{$\true$}{1}}
{\mynode{$\false$}{0}} ]}
\newcommand{\negliteral}[2]{[.{\mynode{$\ite(#1)$}{#2}} {\mynode{$\false$}{0}} {\mynode{$\true$}{1}} ]}

\paragraph{Conditioning}
For a partial valuation $a\in\bool^A$ and a formula $f$ in \gls{dec} it is
possible to compute a formula equivalent to $\instantiate{f}{a}$ also in \gls{dec} as follows:
replace $\ite(v, g, h)$ by $g$ if $v\in A$ and $a(v) = \true$, $h$ if $v\in A$
and $a(v) = \false$ and otherwise leave it as is. Thus, we can
compute $\mc{\instantiate{f}{a}}$ in linear time as well.

\paragraph{Layering} For \femajsat on \gls{dec} formulas, one needs an
extra constraint compared to model counting:
\begin{definition}
  A formula in \gls{dec} is  $(A, X)$-layered if
    $\varsof{f} \subseteq A \uplus X$ (where $\uplus$ denotes
    disjoint union) and
    for any Ite node $\ite(v, f, g)$, we have 
      $
        v \in X \implies
      \varsof{f} \subseteq X
    $.
\end{definition}
This corresponds to Ite nodes on variables in $A$ on top, then those on $X$
below.
Some \gls{dec} compilers like Dsharp~\cite{dsharp} can produce layered
\gls{dec} as it can be used for projected model counting~\cite{projmc}, but
this is significantly more expensive than unconstrained compilation.

\paragraph{Constrained algorithm} We can now solve \femajsat on layered \gls{dec}: 
\begin{definition}[Constrained algorithm~\cite{complan}]
  \label{def_p_emajsat_compute}
  For $f$ in $(A,
  \varset{}\setminus A)$-layered smooth \gls{dec} one defines $C(f)$ and
  $\witness{A}{f}$ as follows:

  \begin{align}
    C(\true) &= 1, \quad C(\false) = 0, \quad \witness{A}{\true} =
    \witness{A}{\false} = a_\bot
    \label{eqn_p_emajsat_true_false} & \\
    (C(\ite(v, g, h))), \witness{A}{\ite(v, g, h)}) &= (C(g) + C(h), a_\bot)
     & \text{when $v\not\in A$}\label{eqn_p_emajsat_uncontrolled_ite} \\
    (C(\ite(v, g, h))), \witness{A}{\ite(v, g, h)}) &=\begin{cases}
      (C(h), \update{\witness{A}{h}}{v}{\false}) & \text{if $C(g) < C(h)$
        } \\
      (C(g),  \update{\witness{A}{g}}{v}{\true}) & \text{otherwise}
        \end{cases}
                                   & \text{when $v\in A$} \label{eqn_p_emajsat_controlled_ite} \\
  \left(C\left(\bigwedge_{i=1}^n g_i\right), \witness{A}{\bigwedge_{i=1}^n g_i}\right) &=
  \left(\prod_{i=1}^n C(g_i), g_1 \concat \dots \concat g_n\right)
    \label{eqn_p_emajsat_and}
  \end{align}
      where $a_\false$ denotes the partial valuation where all variables in $A$
      are mapped to $\false$, and $\update{a}{v}{x}$ denotes the valuation that
      maps $v'$ to $x$ if $v=v'$ else to $a(v')$.
\end{definition}
\begin{proposition}
  \label{prop_p_emajsat_compute}
  $C(f)=\emajsatfun{A}{f}$ and $\witness{A}{f}$ is a witness:
  $\mc{\instantiate{f}{\witness{A}{f}}} = \emajsatfun{A}{f}$.
\end{proposition}

And nodes map to multiplication, chance Ite nodes to addition
and choice Ite nodes to maximum.


To our knowledge this algorithm has no name in the literature, it is mentioned
in \citet{complan,complan+} as a straightforward technique that is not
practical in terms of performance because of constrained compilation, and upon which they intend to improve. We will call this
algorithm \algo{Constrained}.


\paragraph{Unconstrained \femajsat}
If one applies \Cref{def_p_emajsat_compute} on an unconstrained (without
layering constraint) formula, one obtains an upper bound instead:

\begin{definition}[Unconstrained algorithm~\cite{complan}]
  \label{def_p_unconstrained}
  Let $f$ be
  a \gls{dec} formula, not necessarily layered. One defines $N$ inductively as
  follows:
  \begin{align}
  N(\true) &= 1, \quad N(\false) = 0 \label{eqn_p_upper_bound_true_false} & \\
    N(\ite(v, g, h))) &= N(g) + N(h)
     & \text{when $v\not\in A$}\label{eqn_p_upper_bound_uncontrolled_ite} \\
    N(\ite(v, g, h))) &= \max(N(g), N(h))
                                   & \text{when $v\in A$}
                                   \label{eqn_p_upper_bound_controlled_ite} \\
    N\left(\bigwedge_{i=1}^n g_i\right) &= \prod_{i=1}^n N(g_i)
    \label{eqn_p_upper_bound_and}
  \end{align}
\end{definition}

\begin{proposition}
  \label{prop_p_emajsat_unconstrained_upper_bound}
  $N(f) \ge \emajsatfun{A}{f}$.
\end{proposition}

This algorithm was presented in \citet{complan} without name, and we call it
\algo{Unconstrained}. It is still linear in the size
of the formula, and requires a cheaper compilation step.

\paragraph{Complan}
\algo{Complan}~\cite{complan} was designed for Conformant Probabilistic
Planning problems
translated to SSAT~\cite{ssat}: these correspond to SSAT formulas with one quantifier alternation
$\exists{a}\rand{x}f$. It compiles the formula to unconstrained \gls{dec}, and
then explores possible assignments $a$ to choice variables by
a standard branch-and-bound construct based on \algo{Unconstrained}: if $N(a')$
is below the current best value of $a$, then $a'$ can be discarded.

\paragraph{Complan+}\label{par_p_oval}
\algo{Complan+}~\cite{complan+} uses the same structure as \algo{Complan} to
solve \femajsat (for probabilistic planning, or Bayesian inference under the
name \algo{Acemap+}), but replaces the upper bound with a more precise one, which we 
designate as \algo{Oval}. Its principle is quite technical; for our purpose it
suffices to say that it is always more precise than \algo{Unconstrained}, and
that it executes in $O(|f||A|)$ where $|f|$ is the size of the \gls{dec} and
$|A|$ denotes the number choice variables.



As we will see in our experimental
evaluation of \Cref{sec_p_bench}, the cost of constrained compilation makes
algorithms like \algo{Constrained} too expensive for \gls{QRSE}, but upper bounds like \algo{Oval} based on unconstrained compilation are too loose.

\subsection{Our proposition: Relaxation}\label{sec_p_relaxation}

We now propose an algorithm combining the advantages of \algo{Constrained} (precision) and
\algo{Oval} (performance).
We do so by
relaxing the layering constraint on \gls{dec} compilation.
Specifically, we ask for $(A\uplus R, X\setminus R)$-layered \gls{dec}
instead of $(A, X)$-layered previously, with $R$ meant to be small.
This allows the compiler to do decisions on $A\cup R$ instead of just
$A$.
\subsubsection{Upper bound}\label{sec_p_relax_upper}
We adapt \algo{Unconstrained} (\Cref{def_p_unconstrained}) to obtain an upper
bound on $\emajsatfun{A}{f}$.

\begin{definition}[Relaxed upper bound]
  \label{def_p_relaxed_emajsat_upper_bound}
  Let $f$ a formula in $(A\uplus R,
  X)$-layered smooth \gls{dec}. We define $U(f)\in \uint$ inductively as
  follows:
  \begin{align}
    U(\true) &= 1, \quad U(\false) = 0 \label{eqn_p_relaxed_emajsat_true_false} & \\
    U(\ite(v, g, h))) &=   U(g) +
    U(h)  && \text{for $v \in X$}\label{eqn_p_relaxed_emajsat_uncontrolled_ite} \\
    U(\ite(v, g, h)) &=  \max(U(g), U(h))
            && \text{for $v \in A$} \label{eqn_p_relaxed_emajsat_controlled_ite} \\
    U(\ite(v, g, h)) &= U(g) + U(h)
           &&                  \text{for $v \in R$}
                                   \label{eqn_p_relaxed_emajsat_relaxed_ite} \\
    U\left(\bigwedge_{i=1}^n g_i \right) &=  
    \prod_{i=1}^n  U(g_i)
    \label{eqn_p_relaxed_emajsat_and}
  \end{align}
\end{definition}

\begin{figure}
  \begin{center}
\begin{tikzpicture}
\Tree [.{\mynode{$\ite(r)$}{$+$}}
  [.{\mynode{$\wedge$}{$\times$}}
    [.{\mynode{$\ite(a)$}{$\max$}} {\mynode{$\true$}{1}} {\mynode{$\false$}{0}} ]
    [.\node(reused){\mynode{$\ite(x)$}{$+$}}; {\mynode{$\false$}{0}}
      {\mynode{$\true$}{1}} ]
  ]
  [.\node(from){\mynode{$\ite(a)$}{$\max$}} ;
    \edge[draw=none]; {}
    [.{\mynode{$\ite(x)$}{$+$}} {\mynode{$\true$}{1}} {\mynode{$\false$}{0}} ]
  ]
]
\draw (from) -- (reused) ;
\end{tikzpicture}
\end{center}

For $f = \ite(r, a\wedge \neg x, \ite(a, \neg x, x))$,
\Cref{prop_p_relaxed_emajsat_upper_bound} yields $\emajsatfun{\{a\}}{f} \le
\max(1, 0)\times(0+1)+\max(0+1, 1+0) = 2$.
\caption{$(\{a, r\}, \{x\})$-layered \gls{dec} (black), with Relax upper bound
for it (red, \Cref{def_p_relaxed_emajsat_upper_bound}).}
  \label{fig_p_example_formula_relax}
\end{figure}
\begin{proposition}
  \label{prop_p_relaxed_emajsat_upper_bound}
  $U(f) \ge \emajsatfun{A}{f}$.
\end{proposition}
\begin{proof}
  We prove the result by induction on the structure of $f$.
  When we compute $ U(g)$ for $g$ in the lower layer of $f$, only
  \cref{eqn_p_relaxed_emajsat_true_false,eqn_p_relaxed_emajsat_uncontrolled_ite}
  are used. These coincide with computation of $\emajsatfun{A\cup R}{g}$ in
  \Cref{def_p_emajsat_compute}, but
  since $\varsof{g} \cap R = \varnothing$, $ U(g) =\emajsatfun{A\cup R}{g}
  = \emajsatfun{A}{g}$.
  In the case of $ U(f_A)$, where $f_A = \ite(v, g, h)$, $v\in A$, observe that
  \(
    \emajsatfun{A}{f_A} = \max(\emajsatfun{A}{g},\allowbreak \emajsatfun{A}{h})
  \).
  By induction hypothesis, $\emajsatfun{A}{g}  \le  U(g) $ and
$\emajsatfun{A}{h}  \le  U(h) $. As $\max$ is non-decreasing in both its
arguments, we prove the desired result $ \emajsatfun{A}{f_A} \le \max( U(g),  U(h))$.
  Same reasoning works for the product on decomposable And
  nodes.
  The interesting case is the case of a relaxed Ite node: $f_R = \ite(v, g, h)$,
  where $v\in R$ (\cref{eqn_p_relaxed_emajsat_relaxed_ite}). As $v\wedge g$ and $\neg v\wedge h$ have no common model, $\modelsof{f_R} =
  \modelsof{v\wedge g} \uplus \modelsof{\neg v\wedge h}$. Therefore, for a partial
  model $a\in \bool^A$, we have 
  $\mc{\instantiate{f_R}{a}} = \mc{\instantiate{(v\wedge g)}{a}} +
  \mc{\instantiate{(\neg v\wedge h)}{a}} = \mc{\instantiate{g}{a}} + \mc{\instantiate{h}{a}} \le 
  \emajsatfun{A}{g}+\emajsatfun{A}{h}$.
  Hence, $\emajsatfun{A}{f_R} \le
  \emajsatfun{A}{g} + \emajsatfun{A}{h}$. By induction hypothesis
  $\emajsatfun{A}{g} \le U(g)$ and $\emajsatfun{A}{h} \le U(h)$, and thus
  $\emajsatfun{A}{f_R} \le U(h) + U(g)$.
\end{proof}

The principle is the same as in \algo{Unconstrained} except that relaxed Ite nodes map
to addition like chance Ite nodes, whereas during compilation they are in the
upper layer like choice variables. An example is given in
\Cref{fig_p_example_formula_relax}.

\subsubsection{Lower bound}\label{sec_p_relax_lower}
The literature is mostly interested in upper bounds for \femajsat, as they use
it for branch-and-bound algorithms. We use the upper bound as a final
result, so we need a lower bound as well.

With \algo{Constrained}, we compute in linear time a 
witness $\witness{A\cup R}{f}$ for $\emajsatfun{A\cup R}{f}$
(\Cref{def_p_emajsat_compute}): its model count is maximal for
$A\cup R$ in the sense that 
$\mc{\instantiate{f}{\witness{A\cup R}{f}}} = \emajsatfun{f}{A\cup R}$; we can
expect it to have good model count when restricted to $A$.
\begin{definition}[Lower bound]
  \label{def_p_fast_bounds}
  Let $f$ a formula in $(A\uplus R,
  X)$-layered smooth \gls{dec}. Let $w \in \bool^A$ be the partial
  assignment coinciding with $\witness{A\cup R}{f}$ on $A$. We define $L(f) =
  \mc{\instantiate{f}{w}}$.
$L(f) \le \emajsatfun{A}{f}$ by definition of $\emajsatfun{A}{f}$.
\end{definition}


  \subsubsection{Quality of the resulting interval}
  We propose \algo{Relax}, the following algorithm:
  \begin{definition}[Relax]
    \label{def_p_relax}
    For $f$ in \gls{cnf}, a partition of its variables in $A\uplus
    X$, and $R \subseteq X$, first
  compile $f$ to a $(A\uplus R, X \setminus R)$-layered \gls{dec}, then
  compute an interval $[L(f), U(f)]$ for $\emajsatfun{A}{f}$ with
  \Cref{def_p_relaxed_emajsat_upper_bound,def_p_fast_bounds}.
\end{definition}
  The second step is done in linear time in the size of the \gls{dec}.
  The main parameter of \algo{Relax} is $R$ the set of relaxed variables. $R$ is
  meant to be small enough to give good approximation, but large enough to allow
  tractable compilation.
In the limit case where $R$ is empty (no relaxation), the algorithm becomes
identical to \algo{Constrained},
and the resulting interval becomes a singleton.
\begin{proposition}[\algo{Relax} degenerates to \algo{Constrained}]
  \label{prop_relax_degenerate_constrained}
  If $R = \varnothing$, then $U(f)$ and $L(f)$ are
  equal to $\emajsatfun{A}{f}$.
\end{proposition}
\begin{proof}
  In this case, \cref{eqn_p_relaxed_emajsat_relaxed_ite} is not used to compute
  $U$, and the
  other rules computing $U$ are identical to those of \Cref{def_p_emajsat_compute}.
  In \Cref{def_p_fast_bounds}, $w$ is equal to $\witness{A}{f}$ therefore the
  corresponding model count is exactly $\emajsatfun{A}{f}$.
\end{proof}

Conversely, when $R$ contains all of $X$, the algorithm becomes identical to
\algo{Unconstrained}:
\begin{proposition}[\algo{Relax} degenerates to \algo{Unconstrained}]
  \label{prop_relax_degenerate_unconstrained}
  If $R = X$, then $U(f) = N(f)$ where $N$ was defined in
  \Cref{def_p_unconstrained}.
\end{proposition}
\begin{proof}
  In this case, \cref{eqn_p_relaxed_emajsat_uncontrolled_ite} is not used to
  compute $U$, and the
  other rules computing $U$ are identical to those for $N$ in
  \Cref{def_p_unconstrained}, with
  \cref{eqn_p_relaxed_emajsat_relaxed_ite} corresponding to
  \cref{eqn_p_upper_bound_uncontrolled_ite}.
\end{proof}

\begin{theorem}[Precision of \algo{Relax}]
  \label{prop_p_bad_bounds_quality}
  $U(f) \le 2^{\card{R\cap \varsof{f}}}L(f)$
\end{theorem}
\begin{proof}
  The proof involves the intermediate quantity $L'(f) \triangleq
  \emajsatfun{A\cup R}{f}$.
  First we prove that $L(f) \ge L'(f)$.
  Let $w\in \bool^A, w'\in\bool^R$ be defined as $\witness{A\cup R}{f} = w
  \concat w'$.
  Each model $x$ of $\instantiate{f}{\witness{A\cup R}{f}}$ can be mapped to a
  model $w'\concat x$ of $\instantiate{f}{w}$. Therefore,
  $\instantiate{f}{\witness{A\cup R}{f}}$ has fewer models than
  $\instantiate{f}{w}$, which can be written as $L'(f) \le L(f)$.

  Then we prove  $U(f) \le 2^{R\cap\varsof{f}}
  L'(f)$ by induction, comparing rules in
  \Cref{def_p_relaxed_emajsat_upper_bound} and \Cref{def_p_emajsat_compute}.
  For base cases $\true$ and $\false$, $U(f) = L'(f)$.
  For an Ite node with variable in $X$, $ U(f) =  L'(f) = \mc{f}$ and $R\cap
  \varsof{f} = \varnothing$, by layering
  hypothesis.
  In the case of an And node $f = \bigwedge_{i=1}^n g_i$:
  $
     U(f) = \prod_{i=1}^n  U(g_i)
     \hspace{0em plus 1em}\le \prod_{i=1}^n 2^{\card{R\cap\varsof{g_i}}}
               L'(g_i) 
               \hspace{0em plus 1em} =  \prod_{i=1}^n 2^{\card{R\cap \varsof{g_i}}}
               \times \prod_{i=1}^n  L'(g_i)
               =  2^{\sum_{i=1}^n  \card{R\cap \varsof{g_i}}}
               \times L'(f)$
               and observing that $\varsof{f}= \biguplus_{i=1}^n
                 \varsof{g_i}$ we get
               $U(f) = 2^{\card{R\cap\varsof{f}}}
                L'(f)
                $.
  For an Ite node with variable in $A$, \ie $f=\ite(v, g, h)$, $v\in A$:
  $ U(f) = \max( U(g),  U(h)) \le \max( L'(g),
   L'(h)) =  L'(f)$.
  For a relaxed Ite node: $f = \ite(v, g, h)$ with $v\in R$. $U(f) =  U(g)
  +  U(h)$.
  \newcommand{\allbutv}{{2^{\card{R\cap\varsof{f}\setminus\{v\}}}}}
  By induction hypothesis, $ U(g) \le
  {2^{\card{R\cap\varsof{g}}}} L'(g) =
    \allbutv{}  L'(g)$ and similarly for $h$. By summing:
    $
     U(f)
    \le \allbutv \left( L'(g) +  L'(h)\right) 
    \le {2^{\card{R\cap \varsof{f}\setminus\{v\}}}}
    \times 2 \times \max( L'(g)\mathrel{,} L'(h)) 
    = {2^{\card{R\cap\varsof{f}}}}
     \max( L'(g),  L'(h)) 
    \le {2^{\card{R\cap\varsof{f}}}} L'(f)
    $
\end{proof}

\myparagraph{Summary}
\algo{Relax} (\Cref{def_p_relax}) is therefore a parametric algorithm that behaves as
\algo{Constrained} (expensive compilation, exact result) without relaxed
variables, as \algo{Unconstrained} (relatively cheap compilation, loose
approximation) when all chance variables are relaxed, but can also provide a
trade-off between the two: the less relaxed variables there are, the more
precise the answer, but the steeper the computational price.

\section{Implementation \& experiments} \label{sec_p_evaluation}

We first describe our implementations of \femajsat solving (Popcon) and  QRSE (\binsecQRSE),  
then we evaluate the feasibility and relevance of the ideas developed so far.  


\subsection{Popcon, a front-end for \femajsat algorithms} \label{sec_p_tool}
For these experiments we implemented Popcon, a front-end for \femajsat solvers
accepting
SMTLib2(QF\_BV) or DIMACS input. It transparently converts this input to
an appropriate format for the selected algorithm,
including bitblasting with Boolector~\cite{boolector} if necessary, and defers to an existing \femajsat solver or a
reimplementation when not available. Popcon
consists in about 8k lines of Rust.
%
%

\Gls{dec}-based
algorithms (\algo{Oval}, \algo{Constrained}, and
\algo{Complan+}, see \Cref{sec_p_background_femajsat}) are reimplementations,
and compilation is performed by D4~\cite{d4}.
As \algo{Oval} only provides an upper bound, we add the lower bound of
\Cref{sec_p_relax_lower}.

Popcon can also submit the formula to solvers based on different principles:
\algo{dc-ssat}~\cite{dcssat} is a solver for probabilistic planning problems
with arbitrary many SSAT~\cite{ssat} quantifier alternations
(we use a
patched version with a different input format kindly provided by N.-Z. Lee);
\algo{ssatABC}~\cite{ssatABC} is a solver for 2-quantifier SSAT problems based on
clause selection; and \algo{Maxcount}~\cite{maxcount} is an approximate, probabilistic solver for 
\pb{Max\sharpsat}. 
Note that these solvers are not explicitly designed for \femajsat but for more
general problems.


\paragraph{Relaxation}
Popcon provides an implementation of \algo{Relax}
(\Cref{sec_p_relaxation}) by asking D4 for a $(A\uplus R, X)$-layered
\gls{dec} formula instead of a $(A, R\uplus X)$-layered one.
Popcon offers two ways to choose $R$ under the
constraint that $|R| \le r$, where $r$ is a user-controlled parameter: 
\begin{description}
  \item[DFS$(r)$] Starting with $R = \varnothing$, we patch D4 to add
    variables it would have decided if
    not constrained to $R$ until $|R|=r$. $R$
    thus contains the first $r$ variables the compiler wants to decide. D4
    operates in depth-first search order, hence the name; 

  \item[BFS$(r)$] In this mode we mimic the of decisions of model
    counting by running D4 for model counting, and collecting the $r$ top-most
    decided variables in breadth-first-search order in the resulting decision
    tree.
\end{description}



\subsection{\binsecQRSE}\label{sec_p_binsec_qrse}
%
We modified the  binary-level robust symbolic execution engine  BINSEC/RSE~\cite{cavRobust} 
to perform QRSE, using Popcon as a \femajsat solver. As an optimization, Popcon is only used for locations which are reachable (through standard SE queries) but not
robustly reachable (through RSE queries). We also benefit from BINSEC optimizations, such as heavy array preprocessing \cite{farinierArraysMadeSimpler2018}. 
Our tool only supports uniform distributions for uncontrolled
inputs, but it is possible to specify their domain
as intervals and with free-form assumptions. For example, it allows
specifying
\gls{ASLR} for the initial value of the stack register $esp$ as $esp \in
[\code{0xaaaa}, \code{0xbbbb}]$  and
$\textbf{assume} \ esp \% 16 = 0$ (alignment).

\subsection{Experimental evaluation}\label{sec_p_bench}
We consider  the following research questions:
\begin{description}
  \item About quantitative robustness:
\begin{enumerate}[label=\textbf{RQ1.\arabic*}]
  \item\label{rq_p_qrse_precise} Is quantitative robustness more precise than
    reachability and robust reachability in some security  contexts?

  \item\label{rq_p_qrse_merge} Can we find real examples where QRSE does not need path merging, while RSE does? 

  \item\label{rq_p_other_goals}  \citet{cavRobust} argued that
    quantitative approaches would be significantly more expensive than the
    qualitative approach of robust reachability because model counting solvers
    scale worse. Is it the case with QRSE?

\end{enumerate}
\item About \femajsat for QRSE:
\begin{enumerate}[label=\textbf{RQ2.\arabic*}]

  \item Can \femajsat on the formulas coming from \gls{QRSE} be solved exactly
    in practice, and how do the various algorithms we described compare?\label{rq_p_exact}

  \item\label{rq_p_approximate} Can approximate algorithms  solve more
    instances, and at what cost for precision?

  \item\label{rq_p_relax_params} How the number
    of relaxed variables impact  \algo{Relax}?

  \item\label{rq_p_why_bad} Can we venture explanations for the relative poor
    performance of some techniques as shown in \ref{rq_p_exact} and \ref{rq_p_approximate}?
\end{enumerate}
\end{description}

\paragraph{\ref{rq_p_qrse_precise}} We answer this research question with a case
  study about vulnerability-oriented bug triage in the scenario of physical fault injection. 
We consider an attacker which controls part of the input and is able to inject a limited number of faults during 
the program execution. 
The typical question for a security expert is whether a program is vulnerable to such an attacker. 
Reasoning other possible input and faults being extremely complicated for a human, 
this scenario can be partly automated. First,  an automated analysis like \gls{SE} finds possible attack traces, 
\ie one input leading to unexpected behavior, and then these traces are
handed to experts for manual analysis. 

\textit{The practical goal is  to reduce the amount
of manual work needed by limiting the number of  traces sent to the expert, while still discovering all the  
most important attacks.}  


More specifically, we consider  the program
VerifyPIN (specifically, VerifyPIN\_2) from FISSC~\cite{verifypin}, a standard
benchmark from the  physical fault
injection community~\cite{giraudSurveyFaultAttacks2004}.
It is a procedure mimicking a typical password checker 
(ex: PIN entered on an ATM), including security-related
countermeasures. It has two explicit inputs: the 4-byte entered PIN
code (userPIN) and the PIN code stored on the card (cardPIN), 
and returns whether they are equal or not.  
For the sake of 
illustration, we adopt a threat model where the attacker controls the userPIN
only%
\footnote{Other inputs are uncontrolled: the userPIN, but also implicit input, e.g.~uninitialized values accessed due to faults.},  
 and can prevent
the processor from executing one single instruction, effectively replacing it by
 \code{nop} (skip).  
The security question is \textit{``Can such an attacker enter a PIN
distinct from the cardPIN and still be granted access?''}.  
We applied the 126 possible 1-byte and 2-byte wide \code{nop} faults on
VerifyPIN, obtaining 126 \emph{mutants} (i.e., variants of the initial program emulating the considered  faults), and use symbolic execution over them to  
find potential attacks, and distinguish them according to replicability. 
We compare the 4 following approaches experimentally:
    \textbf{SE} the \gls{SE} implementation of BINSEC~\cite{binsec}; 
    \textbf{RSE} the \gls{RSE} implementation of BINSEC/RSE~\cite{cavRobust};  
  \textbf{exact QRSE} our QRSE method with  \algo{Constrained}, the most effective
    exact algorithm in \ref{rq_p_exact};  
   \textbf{relaxed QRSE} our QRSE method with our approximation  \algo{Relax} (best choice according to results in  \ref{rq_p_exact} and \ref{rq_p_approximate}), 
    and to get the best possible answer, we first try
    with $BFS(8)$ for half the timeout (because it provides tight bounds), and if this
    fails, with
    $BFS(128)$ with half the timeout (because it times out least often).

We attempt to identify traces which are above 20\% (highly concerning) or below $10^{-6}$
(noise). For relaxed QRSE, we report traces \textit{provably} in one of the
category above. 
BINSEC and the SMT solver have no timeout, 
 but Popcon is limited to 3 min.
%
%
The
 thresholds mentioned above are chosen to illustrate two approaches: a \textit{conservative} analysis
where only traces with a provably low quantitative robustness are dismissed, and
a more \textit{optimistic} one where one only analyzes traces with
high quantitative robustness.

\begin{table*}
  \centering
  \caption{Comparison of various methods to look for exploitable faults}
  \label{tab_p_comparison_se_qrse}
  \begin{tabular}{c@{\,}c@{\,}c@{\,}ccc}
    \toprule
    \multirow{2}{*}{Method}     & Quantitative & Reported       & \multirow{2}{*}{Time (s)} & \multicolumn{2}{c}{Paths abandoned because of} \\
                                & robustness   & attack traces  & & Z3 UNKNOWN & Popcon timeout              \\
    \midrule
    SE                          & $>0\%$         & 39             & 66                        & 0                                               & --                          \\
    \midrule
    RSE                         & = 100\%      & 0              & 67                        & 25                                              & --                          \\
    \midrule
    \multirow{3}{*}{exact QRSE} & $> 20\%$       & 0              & \multirow{3}{*}{2435}     & \multirow{3}{*}{0}                              & \multirow{3}{*}{13}         \\
                                & $<10^{-6}$   & 23            \\
                                & $\in[10^{-6}, 20\%]$ & 3             \\
    \midrule
    relaxed QRSE                & $> 20\%$       & 2              & \multirow{3}{*}{250}      & \multirow{3}{*}{0}                              & \multirow{3}{*}{\textbf{0}} \\
    BFS(8) then                 & $<10^{-6}$   & 27            \\
    BFS(128)                    & $\in[10^{-6}, 20\%]$ & 10             \\
    \bottomrule
  \end{tabular}
\end{table*}

As shown in \Cref{tab_p_comparison_se_qrse}, \gls{SE} finds 39 attack traces, \gls{RSE}
finds none, and quantitative approaches find an intermediate number of them
depending on the threshold. Exact QRSE has 13 timeouts, but still proves that out of the 39 attacks found
by \gls{SE}, at least 23 are not interesting ($<10^{-6}$).
Relaxed QRSE improves significantly in this regard, as there is no timeout when
using the hybrid BFS(8) then BFS(128) approach. 
%
%
It classifies 27 traces as not interesting, and finds two concerning traces
with quantitative robustness in $[0.992202, 0.992204]$.
Manual analysis on the traces confirms the reported values.
For example, the lowest quantitative robustness (about $2^{-56}$) corresponds to a
mutant where
the attacker must guess 3 bytes of the
cardPIN, the low byte of a register and hope for the top 3 bytes to be zero.
Overall this amounts to
7 bytes, or 56 bits, of luck.
Interestingly, the 6 top faults detected 
are outside the protected code of VerifyPIN,
which proves that the protected part of VerifyPIN admits
no attack with quantitative robustness above
$10^{-4}$ with our threat model.

%
\emph{In the end, this analysis allows to reduce the number of cases to analyze
manually from 39 with standard \gls{SE} to 12 in the conservative scenario
described above, and  2 in the optimistic one -- RSE does not report any case.}
%
%
 Overall, QRSE proves useful here to help focus the attention of the security expert on possibly critical attack traces, 
 and remove noisy ones. 

\paragraph{\ref{rq_p_qrse_merge}} 
We illustrate the benefits of the absence of path merging in a case study about
CVE-2019-20839, a stack buffer overflow in \code{libvncserver}.
The security question is: \textit{Can an attacker controlling the address of the
  server divert control flow to \texttt{0xdeadbeef}?}
 Standard \gls{SE} tells us it is possible for example when the top of the stack 
is at \texttt{0xfff02000} and various other initial conditions are met. But all 
of those, except the arguments, are beyond the control of the attacker, making
this information of little use for vulnerability assessment. 
\gls{RSE} can prove the 
stronger robust reachability: by choosing the right server address, the
attacker can trigger the buffer overflow for all initial conditions. However,
this requires systematic path merging, which is documented to be useful when
used carefully but detrimental to performance when used
systematically~\cite{hansenStateJoiningSplitting2009,kuznetsovEfficientStateMerging2012}. 

As explained in 
\Cref{sec_p_merging}, path merging is not needed in QRSE when only few paths
would need to be merged. Instead, we can attempt to detect single paths with high
quantitative robustness.  
\textit{ On this example, 
QRSE without any path merging is indeed able to find path with quantitative robustness above 30\%. 
The evidence is weaker than full
robust reachability but still a good hint for security.}

\myparagraph{Formula benchmark} To answer the remaining questions about \femajsat, we prepared a benchmark composed of 117
 QRSE-induced \femajsat instances:
%
\textbf{RSE} 92 SMTLib2 formulas obtained by \gls{RSE} on the case
    studies of \citet{cavRobust};
\textbf{VerifyPIN} The 25 distinct SMTLib2 \femajsat problems generated during our case
    study about VerifyPIN (\ref{rq_p_qrse_precise}).
%
The size of these formulas (554 variables and 998 clauses in median after
bitblasting) is comparable to what is found in \citet{ssatABC} (331 variables and 3761
clauses in median).
Problems are run on an Intel Xeon E-2176M CPU (2.70GHz) with a timeout of 20 minutes and memory-out of 2 GB.

\paragraph{\ref{rq_p_other_goals}}
We consider the formula benchmark and compare the following approaches:   \femajsat
  (solved exactly with \algo{Constrained}  or faster but imprecisely with
  \algo{Oval}, the best approaches in \ref{rq_p_exact} and
  \ref{rq_p_approximate}) and \pb{$\forallsymb$SMT} (the quantified version of the formula that RSE has to solve -- we use Z3~\cite{z3}). 
 We also consider the cost of model counting
  \sharpsat (component of \eg probabilistic symbolic
  execution~\cite{geldenhuysProbabilisticSymbolicExecution2012}) and projected
  model counting~\cite{existsSATProjected2015} $\existssymb$\sharpsat (component of \eg quantitative
  information flow~\cite{heusserQuantifyingInformationLeaks2010}), both solved
  with D4~\cite{d4}. 


\begin{figure}
  \begin{minipage}{0.44\textwidth}
\begin{center}
\includegraphics[width=\textwidth]{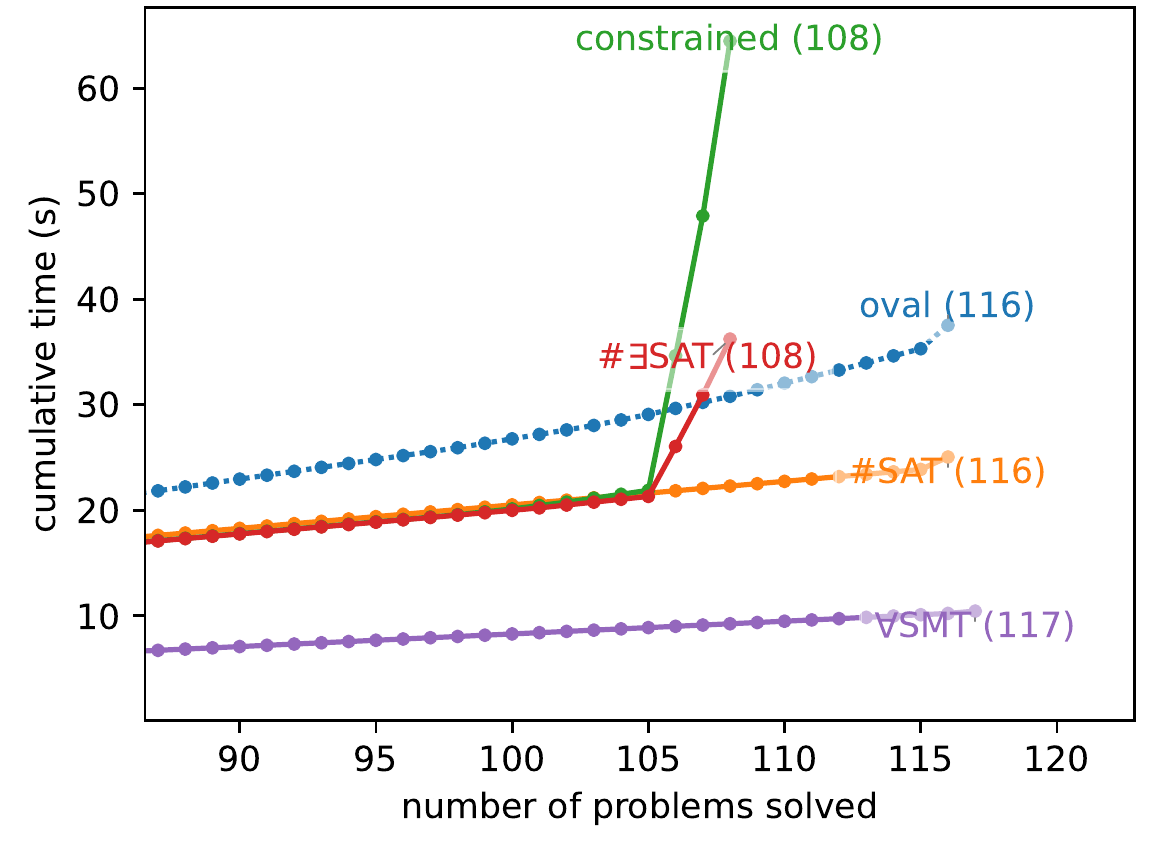}
\end{center}
\captionof{figure}{Comparison of the cost of solving \femajsat to universally quantified
SMT.}
\label{fig_p_popcon_vs_z3}
\end{minipage}
\begin{minipage}{0.48\textwidth}
\begin{center}
\includegraphics[width=\linewidth]{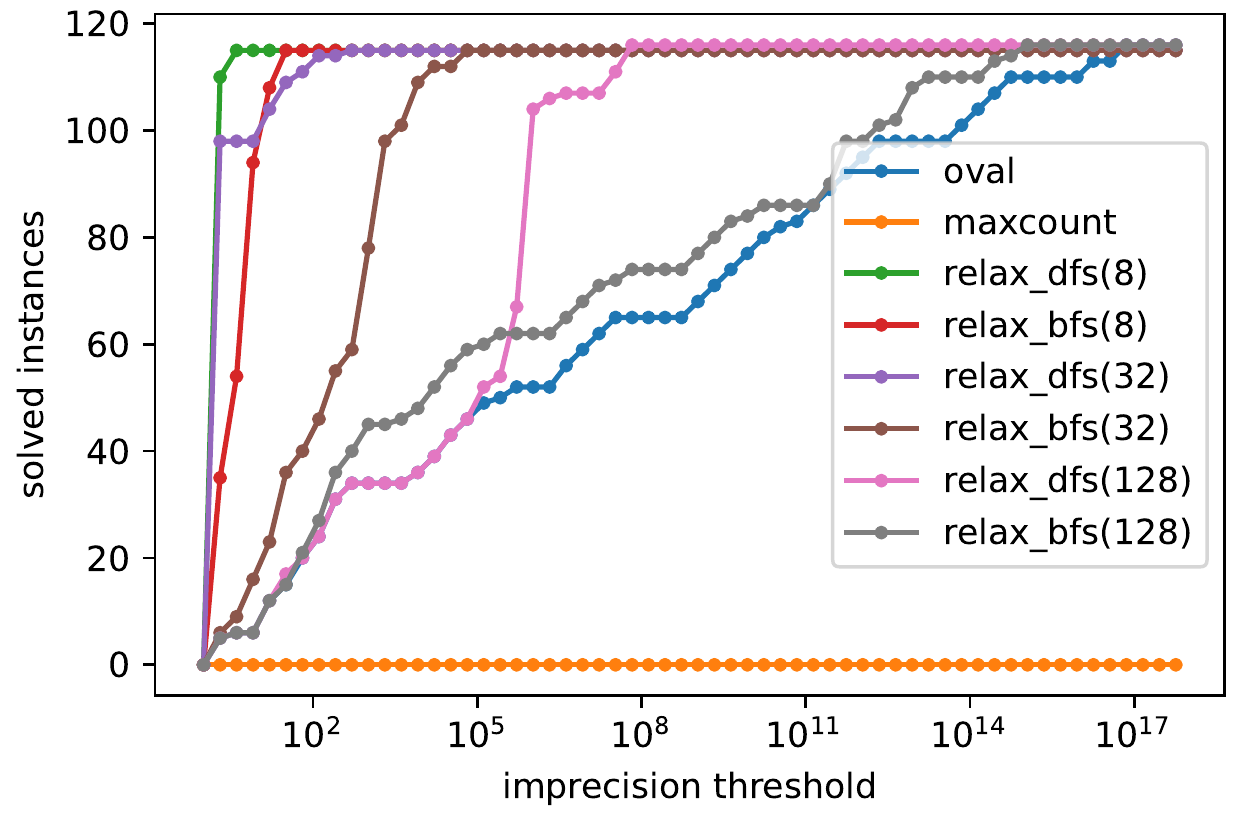}
\end{center}
\caption{Evolution of the number of instances solved under a threshold of
precision by approximate algorithms.}
\label{fig_p_imprecision_vs_solved_instances}
\end{minipage}
\end{figure}

Results are shown in \Cref{fig_p_popcon_vs_z3}. Solving \pb{$\forallsymb{}$SMT} is 7 times faster for 108 instances than
exact \femajsat, and does not suffer from timeouts. \algo{Constrained} times out
9 times, in comparison. Even when completely overlooking the quality of the
result, the inexact algorithm \algo{Oval} is still about 4 times slower, and has
one time-out. \emph{
Quantitative treatment of path constraints generated during (Q)RSE is indeed
significantly more expensive than the corresponding qualitative treatment.}



  \begin{figure}
\begin{center}
\includegraphics[width=0.7\linewidth]{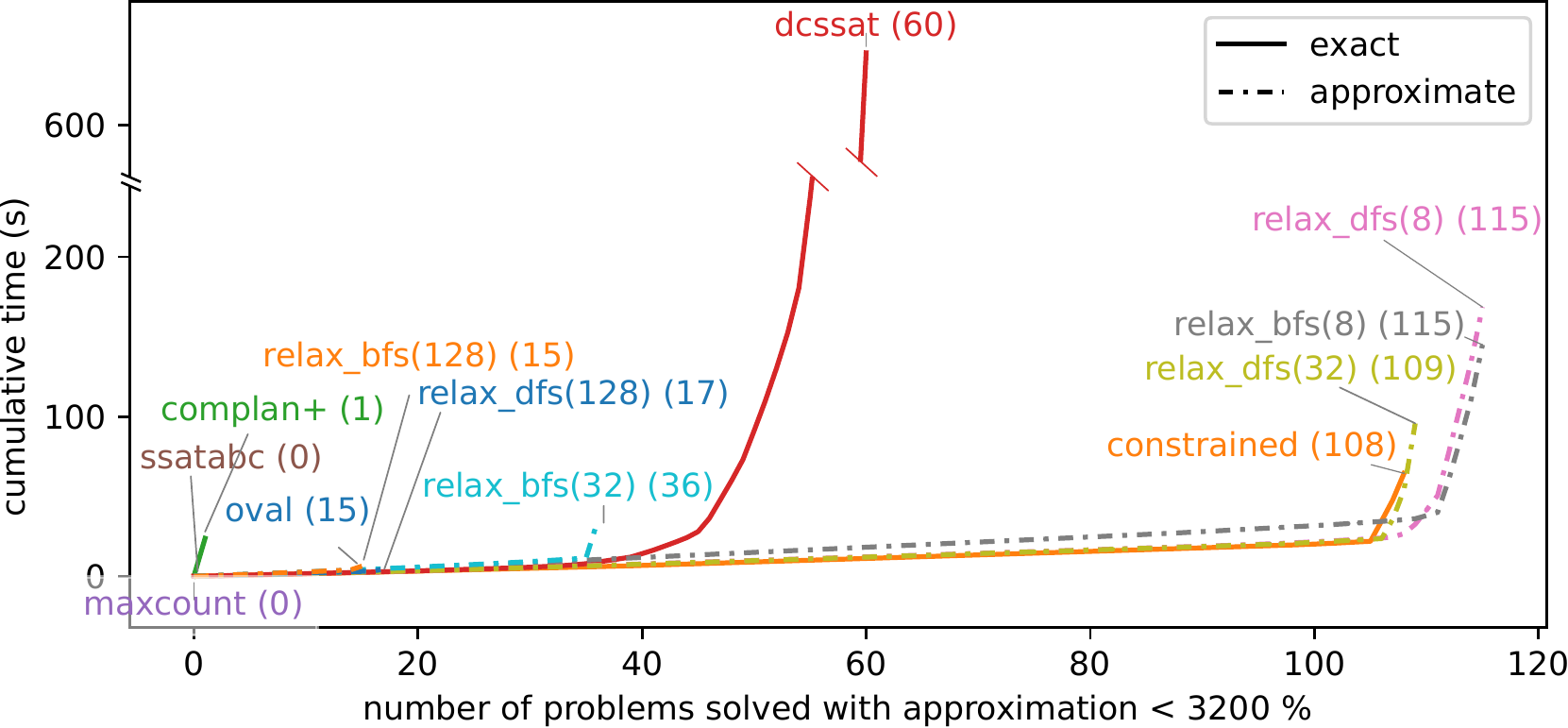}
\end{center}
\caption{Cactus plot of various \femajsat solving algorithms on 117 instances
  coming from QRSE. Dashed lines
  correspond to methods returning an interval  $[l, h]$ rather than an exact
  answer. Only instances solves with imprecision $h/l$ is below 32$\times$ are
  considered solved. The number
of solved instances is given in parentheses.}
\label{fig_p_cactus}
\end{figure}

\paragraph{\ref{rq_p_exact}} Only 
two exact methods can solve a significant number of instances
(\Cref{fig_p_cactus}): \algo{dc-ssat}
(60/117) and \algo{Constrained} (108/117).
This is surprising because \algo{Complan+}
(1/117) was designed to improve over \algo{Constrained},
 as compilation to \gls{dec} is
more expensive when constrained than when unconstrained. This
assumption
is true: \algo{Oval}, which uses unconstrained \gls{dec}
solves 8 more instances than \algo{Constrained} when one ignores the precision of the
result (\Cref{fig_p_imprecision_vs_solved_instances}). The relative poor
performance of \algo{Complan+} therefore comes not from \gls{dec}
compilation but from the branch and
bound step.
Similarly, \algo{ssatABC} solves no instances.

\emph{\algo{Constrained} is the only exact algorithm performing well on formulas generated
by \gls{QRSE} (even better than \algo{Complan+}, which was designed to improve on
it), and it still leaves 7\% of instances unsolved.}

\paragraph{\ref{rq_p_approximate}} 
To solve more than 108/117 instances one needs to resort to approximate
techniques, which return an interval $[l, h]$. \algo{Oval} can
solve 116 instances, and \algo{Relax} can solve from 114 to 116 instances
depending on parameters (\Cref{fig_p_imprecision_vs_solved_instances}). But this is misleading as this ignores the quality of the answer. We call
imprecision  the ratio $h/l$. 
\Cref{fig_p_cactus} shows the number of solved instances under an arbitrary
threshold of 32$\times$, but \Cref{fig_p_imprecision_vs_solved_instances} summarizes
results for other imprecision thresholds.
\algo{Oval} provides
poor approximation, \algo{Relax} can solve 115/117 instances with
imprecision under 4$\times$ with 8 relaxed variables, and
\algo{Maxcount} always times out.

\emph{Approximate algorithms can solve more instances, and
  \algo{Relax} can do so while remaining precise: 115/117 instances
  solved instead
  of 108/117 exactly with imprecision under 4$\times$.}

%
%
%

  \begin{figure}
  \begin{minipage}{0.48\textwidth}
\begin{center}
\includegraphics[width=\columnwidth]{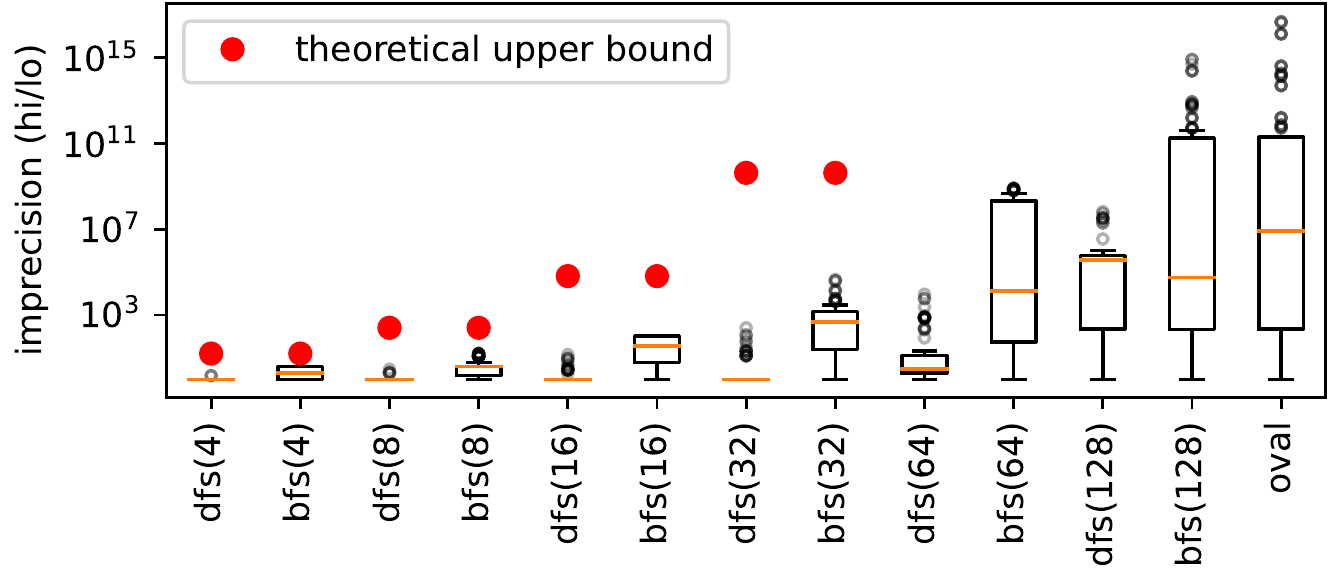}
\end{center}
\scriptsize{Theoretical upper bound (\Cref{prop_p_bad_bounds_quality}) $2^r$ omitted
for $r\ge 64$.}
\captionof{figure}{Box plot of imprecision (upper/lower bound) of approximate \femajsat solving algorithms.}
\label{fig_p_imprecision}
\end{minipage}
\begin{minipage}{0.48\textwidth}
\begin{center}
\includegraphics[width=\columnwidth]{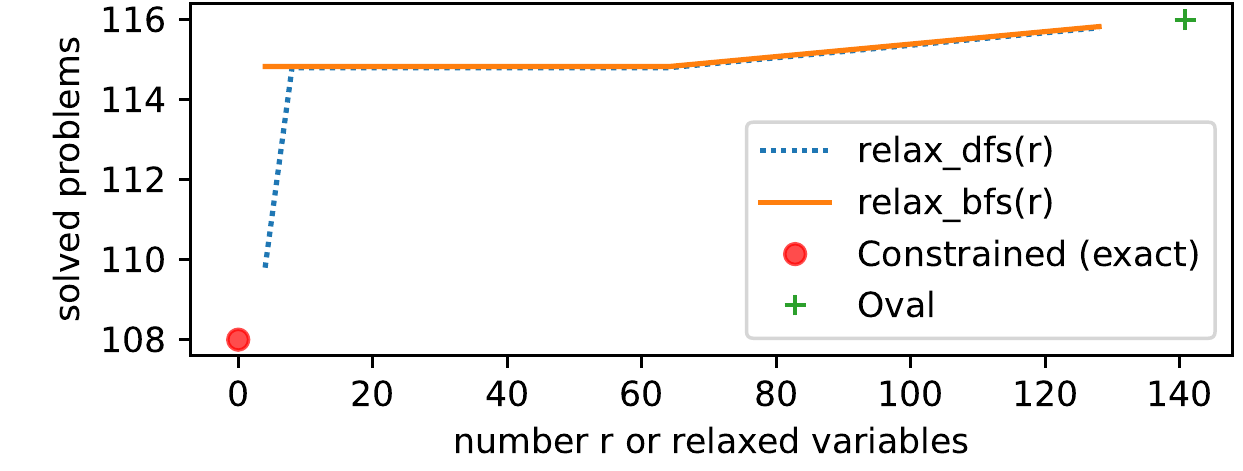}
\end{center}
\captionof{figure}{Solved instances within timeout depending on the number $r$ of relaxation
variables, regardless of precision.}
\label{fig_p_relaxation_solved}
\end{minipage}
\end{figure}

\paragraph{\ref{rq_p_relax_params}} 
The number of instances solved by
\algo{Relax}
within timeout increases with the number $r$ of relaxed variables
  (\Cref{fig_p_relaxation_solved}). Up to 8 more instances can be solved
with relaxation. The imprecision also increases with $r$
(\Cref{fig_p_imprecision}), but it is most often orders of magnitude smaller than the
theoretical bound $2^r$ (\Cref{prop_p_bad_bounds_quality}).  DFS variable order usually yields more precise results,
but for high $r$ values (128) the tendency inverts in median.
As expected
(\Cref{prop_relax_degenerate_unconstrained}), when $r$ becomes large, one obtains
similar behavior as techniques based on fully
unconstrained \gls{dec}, like \algo{Oval}.

\emph{Relaxation can reach a sweet spot between precision and
  efficiency which solves more instances than exact \femajsat with significantly better approximation
  than theoretical bounds.}

\paragraph{\ref{rq_p_why_bad}} Interestingly, replaying our experiments on the test suite of \algo{ssatABC}~\cite{ssatABC} (problems coming for example from probabilistic planning) 
yields radically different results.  \emph{Existing solvers perform better on different kinds of formulas.} More details and experiments in
this direction are available in Supplementary material. 

\section{Related work} 

\paragraph{Quantitative analysis} We attempt at designing a quantitative counterpart to robust
reachability, viewed
as too strict. Such a quantitative relaxation has
already been seen in other domains and is part of a general effort to make formal verification less ``all-or-nothing'':  from
non-interference~\cite{goguenNonInterference} to quantitative information
flow~\cite{heusserQuantifyingInformationLeaks2010},  from traditional model
checking to
probabilistic model
checking~\cite{azizVerifyingContinuousTime1996,hanssonLogicReasoningTime1994}  
or from symbolic execution to probabilistic symbolic execution~\cite{geldenhuysProbabilisticSymbolicExecution2012}. 
These different applications give rise to different counting or probabilistic problems. 
We rely on \femajsat 
while probabilistic verification builds on  standard model
counting~\cite{gomesModelCounting2008}, probabilistic model checking on Markov
chains, and quantitative information flow on  projected model
counting~\cite{existsSATProjected2015}.    


\paragraph{Counting solvers}  Many combinations and extensions
are possible. The branch-and-bound algorithms behind \algo{Complan} and
\algo{Complan+} can be interrupted at any time to obtain a refined, but not
perfect interval. Our algorithm \algo{Relax} could be refined by using bounds
inspired from \algo{Oval} instead of \algo{Unconstrained}, at the price of
significant added complexity. Finally, the choice of the set of relaxed
variables has only been partially explored, and is certainly a direction for
future work.
Some works target model counting beyond propositional
formulas (e.g., for bit-vectors~\cite{searchmc} or integer 
polyhedra~\cite{deloeraEffectiveLatticePoint2004}). That could be a source of inspiration for further developments.


\paragraph{Flakiness} When a branch can be reached robustly, but that outgoing
paths are not robust anymore, then some dependence on uncontrolled input is
introduced. If uncontrolled inputs are taken to be non-deterministic inputs in a
test suite, then this is linked~\cite{cavRobust} to the fact that the test is
\emph{flaky} (has
non-deterministic outcome), which is an
active area of
research~\cite{luoEmpiricalAnalysisFlaky2014,alshammariFlakeFlaggerPredictingFlakiness2021,weiPreemptingFlakyTests2022}.
Quantitative robustness can probably be used to detect further flakiness introduction
locations, in the form of branches which have smaller quantitative robustness
than their parent.

\clearpage


\renewcommand{\exists}{\existssymb}
\renewcommand{\forall}{\forallsymb}
\bibliographystyle{plainnat}
\bibliography{popcon3.bib}

\end{document}